\documentstyle[twocolumn,prd,aps,epsfig]{revtex}
\begin{document}
\twocolumn[\hsize\textwidth\columnwidth\hsize\csname
@twocolumnfalse\endcsname
\title{{\bf 
Strings, loops and others: \\
a critical survey of the present 
approaches to quantum gravity}\\
\vskip.1cm
{\it Plenary lecture on quantum gravity 
at the GR15 conference, Poona, India}}
\author{{\bf Carlo Rovelli}\\
{\it Physics Department, University of Pittsburgh, Pittsburgh, 
PA 15260, USA}\\
rovelli@pitt.edu} 
\date{December 20th, 1997}
\maketitle

\vskip.5cm

\centerline{\bf Abstract}

\begin{abstract} 
	I review the present theoretical attempts to understand the 
	quantum properties of spacetime.  In particular, I illustrate 
	the main achievements and the main difficulties in: string 
	theory, loop quantum gravity, discrete quantum gravity (Regge 
	calculus, dynamical triangulations and simplicial models), 
	Euclidean quantum gravity, perturbative quantum gravity, 
	quantum field theory on curved spacetime, noncommutative 
	geometry, null surfaces, topological quantum field theories 
	and spin foam models.  I also briefly review several recent 
	advances in understanding black hole entropy and attempt a 
	critical discussion of our present understanding of quantum 
	spacetime.
\end{abstract}
\vskip1cm 
\vskip4pc]

\tableofcontents

\newpage
\section{Introduction}

\noindent The landscape of fundamental physics has changed 
substantially during the last one or two decades.  Not long ago, 
our understanding of the weak and strong interactions was very 
confused, while general relativity was almost totally 
disconnected from the rest of physics and was empirically 
supported by little more than its three classical tests.  Then 
two things have happened.  The $SU(3)\times SU(2)\times U(1)$ 
Standard Model has found a dramatic empirical success, showing 
that quantum field theory (QFT) is capable of describing all 
accessible fundamental physics, or at least all non-gravitational 
physics.  At the same time, general relativity (GR) has undergone 
an extraordinary ``renaissance'', finding widespread application 
in astrophysics and cosmology, as well as novel vast experimental 
support -- so that today GR is basic physics needed for 
describing a variety of physical systems we have access to, 
including, as have heard in this conference, advanced 
technological systems \cite{gps}.

These two parallel developments have moved fundamental physics to 
a position in which it has rarely been in the course of its 
history: We have today a group of fundamental laws, the standard 
model and GR, which --even if it cannot be regarded as a 
satisfactory global picture of Nature-- is perhaps the best 
confirmed set of fundamental theories after Newton's universal 
gravitation and Maxwell's electromagnetism.  More importantly, 
there aren't today experimental facts that openly challenge or 
escape this set of fundamental laws.  In this unprecedented state 
of affairs, a large number of theoretical physicists from 
different backgrounds have begun to address {\em the\/} piece of 
the puzzle which is clearly missing: combining the two halves of 
the picture and understanding the quantum properties of the 
gravitational field.  Equivalently, understanding the quantum 
properties of spacetime.  Interest and researches in quantum 
gravity have thus increased sharply in recent years.  And the 
problem of understanding what is a quantum spacetime is today at 
the core of fundamental physics.

The problem is not anymore in the sole hands of the 
relativists.  A large fraction of the particle physicists, 
after having mostly ignored gravity for decades, are now 
exploring issues such as black hole entropy, background 
independence, and the Einstein equations.  Today, in both the 
gr-qc and hep-th sectors of the Los Alamos archives, an 
average of one paper every four is related to quantum 
gravity, a much higher proportion than anytime before.

This sharp increase in interest is accompanied by some results.  
First of all, we do have today some well developed and reasonably 
well defined tentative theories of quantum gravity.  String 
theory and loop quantum gravity are the two major examples.  
Second, within these theories definite physical results have been 
obtained, such as the explicit computation of the ``quanta of 
geometry'' and the derivation of the black hole entropy formula.  
Furthermore, a number of fresh new ideas --for instance,  
noncommutative geometry-- have entered quantum gravity.

A lot of activity does not necessarily mean that the 
solution has been reached, or that it is close.  There is a 
lot of optimism around, but the optimism is not shared by 
everybody.  In particular, in recent years we have 
repeatedly heard, particularly from the string camp, bold 
claims that we now {\em have\/} a convincing and 
comprehensive theory of Nature, including the solution of 
the quantum gravity puzzle.  But many think that these 
claims are not substantiated.  So far, no approach to 
quantum gravity can claim even a single piece of 
experimental evidence.  In science a theory becomes credible 
only after corroborated by experiments -- since then, it is 
just an hypothesis -- and history is full of beautiful 
hypotheses later contradicted by Nature.  The debate between 
those who think that string theory is clearly the correct 
solution and those who dispute this belief is a major 
scientific debate, and one of the most interesting and 
stimulating debates in contemporary science.  This work is 
also meant as a small contribution to this debate.

But if an excess of confidence is, in the opinion of many, far 
premature, a gloomy pessimism, also rather common, is probably 
not a very productive attitude either.  The recent explosion of 
interest in quantum gravity has led to some progress and {\em 
might\/} have taken us much closer to the solution of the puzzle.  
In the last years the main approaches have obtained theoretical 
successes and have produced predictions that are at least 
testable in principle, and whose indirect consequences are being 
explored.

One does not find if one does not search.  The search for 
understanding the deep quantum structure of spacetime, and 
for a conceptual framework within which everything we have 
learned about the physical world in this century could stay 
together consistently, is so fascinating and so 
intellectually important that it is worthwhile pursuing even 
at the risk of further failures.  The research in quantum 
gravity in the last few years has been vibrant, almost in 
fibrillation.  I will do my best to give an overview of what 
is happening.  In the next sections, I present an overview 
of the main present research direction, a discussion of the 
different perspectives in which the problem of quantum 
gravity is perceived by the different communities addressing 
it, and a tentative assessment of the achievements and the 
state of the art.  

I have done my best to reach a balanced view, but the field 
is far from a situation in which consensus has been reached, 
and the best I can offer, of course, is my own biased 
perspective. For a previous overview of the problem of 
quantum gravity, see \cite{IshamOver}. 

\begin{table}[t] 
\twocolumn[\hsize\textwidth\columnwidth\hsize\csname
@twocolumnfalse\endcsname
\begin{tabular}{ccc}
& &   \\
\bf Traditional & \bf Most Popular & \bf New  \\
& &   \\
& &   \\
	\framebox{\ \ 
	   \begin{minipage}{125pt}
	       {\bf Discrete} \\ 
              \mbox{\ \  } Dynamical triangulations\\
              \mbox{\ \  } Regge calculus\\ 
              \mbox{\ \  } Simplicial models\\
              $\rightarrow$ $2^{nd}$ {\em order transition?}
        \end{minipage}} 
& 
    \framebox{\ \ 
        \begin{minipage}{140pt} \mbox{\ } \\
	        {\bf Strings} \\ 
             $\rightarrow$ {\em Black hole entropy} \\ 
         \end{minipage}} 
& 
      \framebox{
         \begin{minipage}{135pt}
              {\bf Noncommutative geometry}	\\
              {\em $\rightarrow$ Quantum theory?}
          \end{minipage}}
\\ && \\
	 \framebox{\ \ 
	      \begin{minipage}{125pt}
	            {\bf Approximate theories} \\
	               \mbox{\ \ } Euclidean quantum gravity \\
	               \mbox{\ \ } Perturbative q.g. \\
                   \mbox{\ \ } $\rightarrow${\em Woodard-Tsamis 
                   effect} \\
	               \mbox{\ \ } QFT on curved spacetime
	       \end{minipage}}
&  
	 \framebox{\ \ 
	       \begin{minipage}{140pt} \mbox{\ } \\ 
	            {\bf Loops} \\  
	             $\rightarrow$ {\em Black hole entropy}\\
	             $\rightarrow$ {\em Eigenvalues of the geometry:}\\ 
	             \\ 
	             \mbox{\ \ } \fbox{ 
	             $ A_{\vec j} = 8 \pi\hbar G 
	             \sum_{i}\sqrt{j_{i}(j_{i}+1)}$}\\
	        \end{minipage}} 
& 
	  \framebox{
         \begin{minipage}{135pt}
              {\bf Null surfaces}	\\
              {\em $\rightarrow$ Observables?}
          \end{minipage}}
\\
& &  \framebox{
           \begin{minipage}{135pt}
              {\bf Spin foam  models}	\\
              {\em $\rightarrow$ convergence of loop, discrete, TQFT 
              and sum-over-histories}
          \end{minipage}} \\ 
      \framebox{
           \begin{minipage}{90pt}{\bf Unorthodox} \\  
                \mbox{\ \ \ }  Sorkin's Posets\\  
                \mbox{\ \ \ }  Finkelstein\\  
                \mbox{\ \ \ }  Twistors\\ 
                \mbox{\ \ \ } \ldots
            \end{minipage}} 
&  &   \\ 
&  &   \\
\end{tabular}
\vskip.5cm
\centerline{\bf Main current approaches to quantum gravity}
\vskip2pc]
\end{table}

\section{Directions}

To get an idea of what the community is working on, I have 
made some amateurish statistical analysis of the subjects of 
the papers in the Los Alamos archives.  The archives which 
are particularly relevant for quantum gravity are gr-qc 
and hep-th.  The split between the two reflects quite 
accurately the two traditions, or the two cultures, that are 
now addressing the problem.  hep-th is almost 3 times larger 
than gr-qc: 295 versus 113 papers per month -- average over 
last year.  In each of the two archives, roughly 1/4 of the 
papers relate to quantum gravity.  Here is a breaking up of 
these paper per field, in an average month:

\vskip1cm
\begin{center}
\begin{tabular}{l c}
	String theory: & 69  \\
	Loop quantum gravity:  & 25  \\
	QFT in curved spaces: & 8  \\
	Lattice approaches: & 7  \\
	Euclidean quantum gravity: & 3  \\
	Non-commutative geometry:\mbox{\ \ } & 3  \\
	Quantum cosmology: & 1  \\
	Twistors: & 1  \\
	Others: & 6  
\end{tabular}
\end{center}
\vskip1cm

Most of the string papers are in hep-th, most of the others  
are in gr-qc.  These data confirm two ideas: that issues 
related to quantum gravity occupy a large part of 
contemporary theoretical research in fundamental physics, 
and that the research is split into two camps.

There are clearly two most popular approaches to quantum 
gravity: a major one, string theory, popular among particle 
physicists, and a (distant) second, loop quantum gravity, 
popular among relativists.  String theory \cite{strings} can 
be seen as the natural outcome of the line of research that 
started with the effort to go beyond the standard model, and 
went through grand unified theories, supersymmetry and 
supergravity.  Loop quantum gravity \cite{loops} can be seen 
as the natural outcome of the line of research that started 
with Dirac's interest in quantizing gravity, which led him 
to the development of the theory of the quantization of 
constrained systems; and continued with the construction of 
canonical general relativity by Dirac himself, Bergmann, 
Arnowit Deser and Misner, the pioneering work in quantum 
gravity of John Wheeler and Brice DeWitt, and the 
developments of this theory by Karel Kuchar, Chris Isham and 
many others.

String theory and loop quantum gravity are characterized by 
surprising similarities (both are based on one-dimensional 
objects), but also by a surprising divergence in philosophy 
and results.  String theory defines a superb ``low'' energy 
theory, but finds difficulties in describing Planck scale 
quantum spacetime directly.  Loop quantum gravity provides a 
beautiful and compelling account of Planck scale quantum 
spacetime, but finds difficulties in connecting to low 
energy physics.

Besides strings and loops, a number of other approaches are being 
investigated.  A substantial amount of energy has been recently 
devoted to the attempt of defining quantum gravity from a 
discretization of general relativity, on the model of lattice QCD 
(Dynamical triangulations, quantum Regge calculus, simplicial 
models).  A number of approaches (Euclidean quantum gravity, old 
perturbative quantum gravity, quantum field theory on curved 
spacetime) aim at describing certain regimes of the quantum 
behavior of the gravitational field in approximate form, without 
the ambition of providing the fundamental theory, even if they 
previously had greater ambitions.  More ``outsider'' 
and radical ideas, such as twistor theory, Finkelstein's 
algebraic approach, Sorkin's poset theory, continue to raise 
interest.  Finally, the last years have seen the appearance of 
radically new ideas, such as noncommutative geometry, the null 
surface formulation, and spin foam models.  I have summarized the 
main approaches in the Table above.

The various approaches are far from independent.  There are 
numerous connections and there is convergence and 
cross-fertilization in ideas, techniques and results. 

\section{Main directions}

\subsection{{\bf String theory}}\label{string}

String theory is by far the research direction which is presently 
most investigated.  I will not say much about the theory, which 
was covered in this conference in the plenary lecture by Gary 
Gibbons.  I will only comment on the relevance of string theory 
for the problem of understanding the quantum properties of 
spacetime.  String theory presently exists at two levels.  First, 
there is a well developed set of techniques that define the 
string perturbation expansion over a given metric background.  
Second, the understanding of the nonperturbative aspects of the 
theory has much increased in recent years \cite{Pol} and in the 
string community there is a widespread faith, supported by 
numerous indications, in the existence of a yet-to-be-found full 
non-perturbative theory, capable of generating the perturbation 
expansion.  There are attempts of constructing this 
non-perturbative theory, generically denoted $M$ theory.  The 
currently popular one is Matrix-theory, of which it is far too 
early to judge the effectiveness \cite{M}.

The claim that string theory solves QG is based on two 
facts.  First, the string perturbation expansion includes 
the graviton.  More precisely, one of the string modes is a 
massless spin two, and helicity $\pm$2, particle.  Such a 
particle necessarily couples to the energy-momentum tensor 
of the rest of the fields \cite{Weinberg} and gives general 
relativity to a first approximation.  Second, the 
perturbation expansion is consistent if the background 
geometry over which the theory is defined satisfies a 
certain consistency condition; this condition turns out to 
be a high energy modification of the Einstein's equations.  
The hope is that such a consistency condition for the 
perturbation expansion will emerge as a full-fledged 
dynamical equation from the yet-to-be-found nonperturbative 
theory.

From the point of view of the problem of quantum gravity, the 
relevant physical results from string theory are two. 
\begin{description}

\item[Black hole entropy.]  The most remarkable physical results 
for quantum gravity is the derivation of the Bekenstein-Hawking 
formula for the entropy of a black hole (See Section \ref{BH}) as 
a function of the horizon area.  This beautiful result has been 
obtained last year by Andy Strominger and Cumrun Vafa 
\cite{stringentropy}, and has then been extended in various 
directions \cite{mal,HorowitzStrominger,DasMathur,%
Maldacena,dongary}.  The result indicates that there is some 
unexpected internal consistency between string theory and QFT on 
curved space.  In section \ref{BH}, I illustrate this result in 
some detail and I compare it with similar results recently 
obtained in other approaches. 

\item[Microstructure of spacetime.]  There are indications that 
in string theory the spacetime continuum is meaningless below the 
Planck length.  An old set of results on very high energy 
scattering amplitudes \cite{amati} indicates that there is no way 
of probing the spacetime geometry at very short distances.  What 
happens is that in order to probe smaller distance one needs 
higher energy, but at high energy the string ``opens up from 
being a particle to being a true string'' which is spread over 
spacetime, and there is no way of focusing a string's collision 
within a small spacetime region.

More recently, in the Matrix-theory nonperturbative formulation 
\cite{M}, the space-time coordinates of the string $x^{i}$ are 
replaced by matrices $(X^{i})^{n}_{m}$.  This can perhaps be 
viewed as a new interpretation of the space-time structure.  The 
continuous space-time manifold emerges only in the long distance 
region, where these matrices are diagonal and commute; while the 
space-time appears to have a noncommutative discretized structure 
in the short distance regime.  This features are still poorly 
understood, but they have intriguing resonances with 
noncommutative geometry \cite{stringnonc} (Section \ref{noncomm}) 
and loop quantum gravity (Section \ref{loop}).

\end{description}

\subsubsection{Difficulties with string theory}

A key difficulty in string theory is the lack of a complete 
nonperturbative formulation.  During the last year, there has 
been excitement for some tentative nonperturbative formulations 
\cite{M}; but it is far too early to understand if these attempts 
will be successful.  Many previously highly acclaimed ideas have 
been rapidly forgotten. 

A distinct and even more serious difficulty of string theory is 
the lack of a {\em background independent\/} formulation of the 
theory.  In the words of Ed Witten:
 \begin{quotation}
	``Finding the right framework for an intrinsic, background 
	independent formulation of string theory is one of the main 
	problems in string theory, and so far has remained out of 
	reach.''  ...  ``This problem is fundamental because it is here 
	that one really has to address the question of what kind of 
	geometrical object the string represents.'' 	\cite{witten}
\end{quotation}
Most of string theory is conceived in terms of a theory 
describing excitations over this or that background, possibly 
with connections between different backgrounds.  This is also 
true for (most) nonperturbative formulations such as Matrix 
theory.  For instance, the (bosonic part of the) lagrangian of 
Matrix-theory that was illustrated in this conference by Gary 
Gibbons is
\begin{equation}
	L \sim \frac{1}{2}{\rm Tr}\left(\dot X^{2} + 
	\frac{1}{2} [X^{i},X^{j}]^{2}
	\right). 
\end{equation}
The indices that label the matrices $X^{i}$ are raised and 
lowered with a Minkowski metric, and the theory is Lorentz 
invariant.  In other words, the lagrangian is really 
\begin{equation}
L \sim \frac{1}{2}{\rm Tr}\left(g^{00}g^{ij}\dot X_{i}\dot X_{j} 
+ \frac{1}{2} g^{ik}g^{jl}[X_{i},X_{j}][X_{k},X_{l}]\right),
\end{equation}
where $g$ is the flat metric of the background.  This shows 
that there is a non-dynamical metric, and an implicit flat 
background in the action of the theory.  (For attempts to explore 
background independent Matrix-theory, see \cite{smo}).

But the world is not formed by a fixed background over which 
things happen.  The background itself is dynamical.  In 
particular, for instance, the theory should contain quantum 
states that are quantum superpositions of different backgrounds 
-- and presumably these states play an essential role in the deep 
quantum gravitational regime, namely in situations such as the 
big bang or the final phase of black hole evaporation.  The 
absence of a fixed background in nature (or active diffeomorphism 
invariance) is the key general lessons we have learned from 
gravitational theories.  I discuss this issue in more detail in 
section \ref{discussion}.  In the opinion of many, until string 
theory finds a genuine background independent formulations, it 
will never have a convincing solution of the quantum gravity 
puzzle.  Until string theory describes excitations located over a 
metric background, the central problem of a true merge of general 
relativity and quantum mechanics, and of understanding quantum 
spacetime, has not been addressed.

Finally there isn't any direct or indirect experimental support 
for string theory (as for any other approach to quantum gravity).  
Claiming, as it is sometimes done, that a successful physical prediction of 
string theory is GR is a nonsense for various reasons.  First, by 
the same token one could claim that the $SU(5)$ grand unified 
theory (an extremely beautiful theoretical idea, sadly falsified 
by the proton decay experiments) is confirmed by the fact that it 
predicts electromagnetism.  Second, GR did not emerge as a 
surprise from string theory: it is because string theory could describe 
gravity that it was taken seriously as a unified theory.  Third, 
if GR was not known, nobody would have thought of replacing the 
flat spacetime metric in the string action with a curved and 
dynamical metric.  ``Predicting'' a spin-two particle is no big 
deal in a theory that predicts any sort of still unobserved other 
particles.  The fact that string theory includes GR is a 
necessary condition for taking it seriously as a promising 
tentative theory of quantum gravity, not an argument in support 
of its physical correctness.

An important remark is due in this regard.  All the testable 
predictions made after the standard model, such as proton decay, 
monopols, existence of supersymmetric partners, exotic particles 
\ldots, have, so far, {\em all\/} failed to be confirmed by time 
and money consuming experiments designed to confirm them.  The 
comparison between these failed predictions and the extraordinary 
confirmation obtained by {\em all\/} the predictions of the 
standard model (neutral currents, W and Z particles, top quark 
...)  may contain a lesson that should perhaps make us reflect.  
If all predictions are confirmed until a point, and all 
predictions fail to be confirmed afterwards, one might suspect 
that a wrong turn might have been taken at that point.  Contrary 
to what sometimes claimed, the theoretical developments that 
have followed the standard model, such as for instance 
supersymmetry, are {\em only\/} fascinating but non-confirmed 
{\em hypotheses\/}.  As far we really know, nature may very well 
have chosen otherwise.

Experimental observation of supersymmetry might very well change 
this balance, and may be in close reach.  But we have been 
thinking that observation of supersymmetry was around the corner 
for quite sometime now, and it doesn't seem to show up yet.  
Until it does, if there is any indication at all coming from the 
experiments, this indication is that all the marvelous ideas that 
followed the standard model may very well be all in the wrong 
direction.  The great tragedy of science, said TH Huxley, is the 
slaying of a beautiful hypothesis by brute facts.

In spite of these difficulties, string theory is today, without 
doubt, the leading and most promising candidate for a quantum 
theory of gravity.  It is the theory most studied, most developed 
and closer to a comprehensive and consistent framework.  It is 
certainly extremely beautiful, and the recent derivation of the 
black hole entropy formula with the exact Bekenstein-Hawking 
coefficient represents a definite success, showing that the 
understanding of the theory is still growing.

\subsubsection{String cosmology}

There has been a burst of recent activity in an outgrowth of 
string theory denoted string cosmology \cite{Veneziano}. The 
aim of string cosmology is to extract physical consequences from 
string theory by applying it to the big bang.  The idea is to 
start from a Minkowski flat universe; show that this is unstable 
and therefore will run away from the flat (false-vacuum) state.  
The evolution then leads to a cosmological model that starts off 
in an inflationary phase.  This scenario is described using 
minisuperspace technology, in the context of the low energy 
theory that emerge as limit of string theory.  Thus, first one 
freezes all the massive modes of the string, then one freezes all 
massless modes except the zero modes (the spatially constant 
ones), obtaining a finite dimensional theory, which can be 
quantized non-perturbatively.  The approach has a puzzling 
aspect, and a very attractive aspect.

The puzzling aspect is its overall philosophy.  Flat space is 
nothing more than an accidental local configuration of the 
gravitational field.  The universe as a whole has no particular 
sympathy for flat spacetime.  Why should we consider a 
cosmological model that begins with a flat spacetime?  To make 
this point clear using a historical analogy, string cosmology is 
a little bit as if after Copernicus discovered that the Earth is 
not in the center of the solar system, somebody would propose the 
following explanation of the {\em birth\/} of the solar system: 
at the beginning the Earth was in the center.  But this 
configuration is unstable (it is!), and therefore it decayed into 
another configuration in which the Earth rotates around the Sun.  
This discussion emphasizes the profound cultural divide between 
the relativity and the particle physicists' community, in dealing 
with quantum spacetime. 

The compelling aspect of string cosmology, on the other hand, is 
that it provides a concrete physical application of string 
theory, which might lead to consequences that are in principle 
observable.  The spacetime emerging from the string cosmology 
evolution is filled with a background of gravitational waves 
whose spectrum is constrained by the theory.  It is not 
impossible that we will be able to measure the gravitational wave 
background not too far in the future, and the prospect of having 
a way for empirically testing a quantum gravity theory is very  
intriguing.

\subsection{{\bf Loop quantum gravity}}\label{loop}

The second most popular approach to quantum gravity, and the most 
popular among relativists is loop quantum gravity \cite{loops}.  
Loop quantum gravity is presently the best developed alternative 
to string theory.  Like strings, it is not far from a complete 
and consistent theory and it yields a corpus of definite physical 
predictions, testable in principle, on quantum spacetime.

Loop quantum gravity, however, attacks the problem from the 
opposite direction than string theory.  It is a non-perturbative 
and background independent theory to start with.  In other words, 
it is deeply rooted into the conceptual revolution generated by 
general relativity.  In fact, successes and problems of loop 
quantum gravity are complementary to successes and problems of 
strings.  Loop quantum gravity is successful in providing a 
consistent mathematical and physical picture of non perturbative 
quantum spacetime; but the connection to the low energy dynamics 
is not yet completely clear.

The general idea on which loop quantum gravity is based is the 
following.  The core of quantum mechanics is not identified with 
the structure of (conventional) QFT, because conventional QFT 
presupposes a background {\em metric\/} spacetime, and is 
therefore immediately in conflict with GR. Rather, it is 
identified with the general structure common to all quantum 
systems.  The core of GR is identified with the absence of a 
fixed observable background spacetime structure, namely with 
active diffeomorphism invariance.  Loop quantum gravity is thus a 
quantum theory in the conventional sense: a Hilbert space and a 
set of quantum (field) operators, with the requirement that its 
classical limit is GR with its conventional matter couplings.  
But it is not a QFT over a metric manifold.  Rather, it is a 
``quantum field theory on a differentiable manifold'', respecting 
the manifold's invariances and where only coordinate independent 
quantities are physical.

Technically, loop quantum gravity is based on two inputs: 
\begin{itemize} 
\item The formulation of classical GR based on the Ashtekar 
connection \cite{ashtekar}.  The version of the connection 
now most popular is not the original complex one, but an 
evolution of the same, in which the connection is real.  
\item The choice of the holonomies of this connection, 
denoted ``loop variables'', as basic variables for the 
quantum gravitational field \cite{rovellismolin}.
\end{itemize} 
This second choice determines the peculiar kind of quantum theory 
being built.  Physically, it corresponds to the assumption that 
excitations with support on a loop are normalizable states.  This 
is the key technical assumption on which everything relies.

It is important to notice that this assumption fails in 
conventional 4d Yang Mills theory, because loop-like excitations 
on a metric manifold are too singular: the field needs to be 
smeared in more dimensions Equivalently, the linear closure of 
the loop states is a ``far too big'' non-separable state space.  
This fact is the major source of some particle physicists's 
suspicion at loop quantum gravity.  What makes GR different from 
4d Yang Mills theory, however, is nonperturbative diffeomorphism 
invariance.  The gauge invariant states, in fact, are not 
localized at all -- they are, pictorially speaking, smeared by 
the (gauge) diffeomorphism group all over the coordinates 
manifold.  More precisely, factoring away the diffeomorphism 
group takes us down from the state space of the loop excitations, 
which is ``too big'', to a separable physical state space of the 
right size \cite{rovellidepietri,zapata}.  Thus, the consistency 
of the loop construction relies heavily on diffeomorphism 
invariance.  In other words, the diff-invariant invariant loop 
states (more precisely, the diff-invariant spin network states) 
are not physical excitations of a field {\em on\/} spacetime.  
They are excitations {\em of \/} spacetime itself.

Loop quantum gravity is today ten years old.  Actually, the first 
announcement of the theory was made in India precisely 10 years 
ago, {\em today\/}!  \cite{india87} \ \   In the last years, the 
theory has grown substantially in various directions, and has 
produced a number of results, which I now briefly illustrate.

\begin{description}

\item[Definition of theory.]  The mathematical structure of the 
theory has been put on a very solid basis.  Early difficulties 
have been overcome.  In particular, there were three major 
problems in the theory: the lack of a well defined scalar 
product, the overcompleteness of the loop basis, and the 
difficulty of treating the reality conditions.
\begin{itemize}
\item The problem of the lack of a scalar product on the Hilbert 
space has been solved with the definition of a diffeomorphism  
invariant measure on a space of connections 
\cite{measure}.  Later, it has also became clear that the same 
scalar product can be defined in a purely algebraic manner 
\cite{rovellidepietri}.  The state space of the theory is 
therefore a genuine Hilbert space $\cal H$.  \item The 
overcompleteness of the loop basis has been solved by the 
introduction of the spin network states \cite{spinnet}.  A spin 
network is a graph carrying labels (related to $SU(2)$ 
representations and called ``colors'') on its links and its 
nodes.

\begin{figure} 
\centerline{\mbox{\epsfig{file=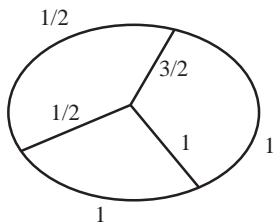}}} 
\caption{Figure 1: A simple spin network. Only the 
coloring of the links is indicated.}
\end{figure}

Each spin network defines a spin network state, and 
the spin network states form a (genuine, non-overcomplete) 
orthonormal basis in $\cal H$. 
 \item The difficulties with the reality conditions have been 
circumvented by the use of the real formulation 
\cite{barbero,thiemann}. 
\end{itemize}
The kinematics of loop quantum gravity is now defined with a 
level of rigor characteristic of mathematical physics 
\cite{mathfound} and the theory can be defined using various 
alternative techniques \cite{rovellidepietri,depietri}.

\item[Hamiltonian constraint.]  A rigorous definition version of 
the hamiltonian constraint equation has been constructed 
\cite{thiemannham}.  This is anomaly free, in the sense that the 
constraints algebra closes (but see later on).  The hamiltonian 
has the crucial properties of acting on nodes only, which implies 
that its action is naturally discrete and combinatorial 
\cite{rovellismolin,rovelliham}.  This fact is at the roots of 
the existence of exact solutions \cite{rovellismolin,solutions}, 
and of the possible finiteness of the theory

\item[Matter.]  The old hope that QFT divergences could be 
cured by QG has recently received an interesting 
corroboration.  The matter part of the hamiltonian 
constraint is well-defined without need of renormalization 
\cite{thiemannmatt}.  Thus, a main possible stumbling block 
is over: infinities did not appear in a place where they 
could very well have appeared. 

\item[Physical Results.] \mbox{\ }\\ 

\begin{description}
\item[Black hole entropy.]  The first important physical result 
in loop quantum gravity is a computation of black hole entropy   
\cite{loopentropy1,loopentropy2,loopentropy3}.  I describe this 
result and I compare it with other derivations in section \ref{BH}. 

\item[Quanta of geometry.]  A very exciting development in 
quantum gravity in the last years has been by the 
computations of the quanta of geometry.  That is, the 
computation of the discrete eigenvalues of area and volume.  
I describe this result a bit more in detail in the next 
section.

\end{description}
\end{description}

\subsubsection{Quanta of Geometry}\label{quanta}

In quantum gravity, any quantity that depends on the metric 
becomes an operator.  In particular, so do the area $A$ of a 
given (physically defined) surface, or the volume $V$ of a given 
(physically defined) spatial region.  In loop quantum gravity, 
these operators can be written explicitly.  They are 
mathematically well defined self-adjoint operators in the Hilbert 
space $\cal H$.  We know from quantum mechanics that certain 
physical quantities are quantized, and that we can compute their 
discrete values by computing the eigenvalues of the corresponding 
operator.  Therefore, if we can compute the eigenvalues of the 
area and volume operators, we have a physical prediction on the 
possible quantized values that these quantities can take, at the 
Planck scale.  These eigenvalues have been computed in loop 
quantum gravity \cite{discreteness}.  Here is for instance the 
main sequence of the spectrum of the area
\begin{equation}
  A_{\vec j} = 8\pi\gamma\, \hbar G\, \sum_{i}\sqrt{j_{i}(j_{i}+1)}. 
  \label{area}
\end{equation}
$\vec j=(j_{1},\ldots,j_{n})$ is an $n$-tuplet of 
half-integers, labeling the eigenvalues, $G$ and $\hbar$ are 
the Newton and Planck constants, and $\gamma$ is a 
dimensionless free parameter, denoted the Immirzi parameter, 
not determined by the theory \cite{Immirzi,RT}.  A similar 
result holds for the volume. The spectrum (\ref{area}) 
has been rederived and completed using various different 
techniques \cite{rovellidepietri,abhayarea,altriarea}.  

These spectra represent solid results of loop quantum 
gravity.  Under certain additional assumptions on the 
behavior of area and volume operators in the presence of 
matter, these results can be interpreted as a corpus of 
detailed quantitative predictions on hypothetical Planck 
scale observations.  Eq.(\ref{area}) plays a key role also 
in black hole physics; see Section \ref{BH}.

Besides its direct relevance, the quantization of the area 
and thee volume is of interest because it provides a 
physical picture of quantum spacetime.  The states of the 
spin network basis are eigenstates of some area and volume 
operators.  We can say that a spin network carries quanta of 
area along its links, and quanta of volume at its nodes.  
The magnitude of these quanta is determined by the coloring.  
For instance, the half-integers $j_{1}\ldots j_{n}$ in 
(\ref{area}) are the coloring of the spin network's links 
that cross the given surface.  Thus, a quantum spacetime can 
be decomposed in a basis of states that can be visualized as 
made by quanta of volume (the intersections) separated by 
quanta of area (the links).  More precisely, we can view a 
spin network as sitting on the {\em dual\/} of a cellular 
decomposition of physical space.  The nodes of the spin 
network sit in the center of the 3-cells, and their coloring 
determines the (quantized) 3-cell's volume.  The links of 
the spin network cut the faces of the cellular decomposition, 
and their color $\vec j$ determine the (quantized) areas of 
these faces via equation (\ref{area}).  See Figure 2.

\begin{figure} 
\centerline{\mbox{\epsfig{file=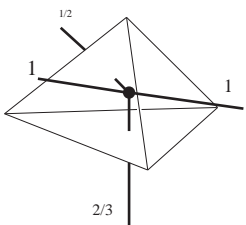}}} 
\caption{Figure 2: A node of a spin network (in bold) and its dual 
3-cell (here a tetrahedron).  The coloring of the node 
determines the quantized volume of the tetrahedron.  The 
coloring of the links (shown in the picture) determines the 
quantized area of the faces via equation (\ref{area}).  Here 
the vector $\vec j$ has a single component, because each 
face is crossed by one link only.}
\end{figure}

Finally, a recent evolution in loop quantum gravity looks 
particularly promising.  A spacetime, path integral-like, 
formulation theory has been derived from the canonical 
theory.  This evolution represents the merging between 
loop quantum gravity and other research directions; I 
illustrate it in section \ref{foam}.

\subsubsection{Difficulties with loop quantum gravity} 

While the kinematics of quantum spacetime is well understood, its 
dynamics is much less clear.  The main problem originates from 
the quantum constraint algebra.  The algebra is anomaly free, in 
the sense that it closes.  However, it differs from the classical 
one in a subtle sense \cite{anomaly}.  For this reason and others 
\cite{variants}, doubts have been raised on the correctness of 
the proposed form of the hamiltonian constraint, and  
variants have been considered \cite{rr,variants}.  A solid proof 
that any of these versions yields classical GR in the classical 
limit, however, is lacking.

Furthermore, a systematic way of extracting physical prediction 
from the theory, analogous, say, to the perturbative QFT scattering 
expansion, is not yet available.  Finally, a description of the 
Minkowski vacuum state is notably absent.  Thus, the theory 
describe effectively quantum spacetime, but the extent to which 
low energy physics can be recovered is unclear.

In summary, the mathematics of the theory is solidly defined and 
understood from alternative points of view.  Longstanding 
problems (lack of a scalar product, overcompleteness of the loop 
basis and reality condition) have been solved.  This kinematics 
provides a compelling description of quantum spacetime in terms 
of discrete excitations of the geometry carrying discretized 
quanta of area and volume.  The theory can be extended to include 
matter, and there are strong indications that ultraviolet 
divergences do not appear.  A spacetime covariant version of the 
theory, in the form of a topological sum over surfaces is under 
development.  The main physical results derived so far are the 
computation of the eigenvalues of area and volume, and the 
derivation of the black hole entropy formula.  Version of the 
dynamics exists, but a proof that the classical limit is 
classical GR is still lacking.  The main open problems are to 
determine the correct version of the hamiltonian constraint and 
to understand how to describe the low energy regime.

\section{Traditional approaches}

\subsection{{\bf Discrete approaches}} 
\label{discrete}

Discrete quantum gravity is the program of regularizing classical 
GR in terms of some lattice theory, quantize this lattice theory, 
and then study an appropriate continuum limit, as one may do in 
QCD. There are three main ways of discretizing GR.

\subsubsection{Regge calculus}

Regge introduced the idea of triangulating spacetime by means of a 
simplicial complex and using the lengths $l_{i}$ of the links of 
the complex as gravitational variables \cite{Regge}.  The theory 
can then be quantized by integrating over the lengths $l_{i}$ of 
the links.  For a recent review and extensive references see 
\cite{ReggeCalculus}.  Recent work has focused in problems such 
as the geometry of Regge superspace \cite{WilliamsHartle} and 
choice of the integration measure \cite{MenottiPeirano}.  Some 
difficulties of this approach have recently been discussed in 
\cite{Reggedif}, where it is claimed that quantum Regge 
calculus fails to reproduce the results obtained in the continuum 
in the lower dimensional cases where the continuum theory is 
known.

\subsubsection{Dynamical triangulations}

Alternatively, one can keep the length of the links fixed, and 
capture the geometry by means of the way in which the simplices 
are glued together, namely by the triangulation.  The 
Einstein-Hilbert action of Euclidean gravity is approximated by a 
simple function of the total number of simplices and links, and 
the theory can be quantized summing over distinct triangulations.  
For a detailed introduction, a recent review, and all relevant 
references [and, last but not least, Mauro Carfora's (and 
Gaia's!)  drawings], see \cite{carfora}.  There are two coupling 
constants in the theory, roughly corresponding to the Newton and 
cosmological constants.  These define a two dimensional space of 
theories.  The theory has a nontrivial continuum limit if in this 
parameter space there is a critical point corresponding to a 
second order phase transition.  The theory has phase transition 
and a critical point \cite{AmbjornJurkiewicz}.  The transition 
separates a phase with crumpled spacetimes from a phase with 
``elongated'' spaces which are effectively two-dimensional, with 
characteristic of a branched polymer \cite{BakkerSmit,AmbJ}.  
This polymer structure is surprisingly the same as the one that 
emerges from loop quantum gravity at short scale.  Near the 
transition, the model appears to produce ``classical'' $S^4$ 
spacetimes, and there is evidence for scaling, suggesting a 
continuum behavior \cite{BakkerSmit}.  However, evidence has been 
contradictory on whether of not the critical point is of the 
second order, as required for a nontrivial scaling limit.  The 
consensus seems to be clustering for a first order transition 
\cite{Bialas}.  This could indicate that the approach does not 
lead to a continuum theory.

Ways out from this serious impasse are possible.  First, 
it has been suggested that even a first order phase transition 
may work in this context \cite{Smit}.  Second, 
Br\"ugmann and Marinari have noticed that there is some freedom 
in the definition of the measure in the sum over triangulations, 
and have suggested (before the transition was shown to be first 
order) that taking this into account might change the nature of 
the transition \cite{BrugmanMarinari}.

\subsubsection{Ponzano-Regge state sum models}\label{pr}

A third road for discretizing GR was opened by a celebrated paper 
by Ponzano and Regge \cite{PonzanoRegge}.  Ponzano and Regge 
started from a Regge discretization of three-dimensional GR and 
introduced a second discretization, by posing the ansatz 
(the {\em Ponzano Regge ansatz\/}) that the lengths $l$ assigned to 
the links are discretized as well, in half-integers in Planck 
units 
\begin{equation}
	l=\hbar G\, j , \ \ \ \ \ j=0,\frac{1}{2},1,\ldots 
	\label{PonzanoRegge}
\end{equation}
(Planck length is $\hbar G$ in 3d.)  The half integers $j$ 
associated to the links are denoted ``coloring'' of the 
triangulation.  Coloring can be viewed as the assignment of a 
$SU(2)$ irreducible representation to each link of the Regge 
triangulation.  The elementary cells of the triangulation are 
tetrahedra, which have six links, colored with six $SU(2)$ 
representations.  $SU(2)$ representation theory naturally assigns 
a number to a sextuplet of representations: the Wigner 6-$j$ 
symbol.  Rather magically, the product over all tetrahedra of 
these 6-$j$ symbols converges to (the real part of the exponent 
of) the Einstein Hilbert action.  Thus, Ponzano and Regge were 
led to propose a quantization of 3d GR based on the partition 
function
\begin{equation}
Z \sim \!\! \sum_{coloring}\ \prod_{tetrahedra}\!\! \mbox{\rm 6-}j({\rm 
color\ of\ the\  tetrahedron})
	\label{PR}
\end{equation}
(I have neglected some coefficients for simplicity).  They also 
provided arguments indicating that this sum is independent from 
the triangulation of the manifold.  

The formula (\ref{PR}) is simple and beautiful, and the idea has 
recently had many surprising and interesting developments.  
Three-dimensional GR was quantized as a topological field theory 
(see Section \ref{tqft}) in \cite{Witten3dGR} and using loop 
quantum gravity in \cite{loop3dGR}.  The Ponzano-Regge 
quantization based on equation (\ref{PR}) was shown to be 
essentially equivalent to the TQFT quantization in \cite{Ooguri}, 
and to the loop quantum gravity in \cite{RovelliPonzano}. (For an 
extensive discussion of quantum gravity in 3 dimensions and what 
we have learned from it, see \cite{Carlip3}.)

Something remarkable happens in establishing the relation between 
the Ponzano-Regge approach and the loop approach: the 
Ponzano-Regge ansatz (\ref{PonzanoRegge}) can be {\em derived\/} 
from loop quantum gravity \cite{RovelliPonzano}.  Indeed, 
(\ref{PonzanoRegge}) turns out to be nothing but the 2d  
 version of the 3d formula 
(\ref{area}), which gives the quantization of the area.  
Therefore, a key result of quantum gravity of the last years, 
namely the quantization of the geometry, derived in the loop 
formalism from a full fledged nonperturbative quantization of GR, 
was  anticipated as an ansatz by the intuition of 
Ponzano and Regge. 

Surprises continued with the attempts to extend these ideas to 4 
dimensions.  These attempts have lead to a fascinating 
convergence of discrete gravity, topological quantum field 
theory, and loop quantum gravity.  I describe these developments 
in section \ref{foam}.  I only notice here that the connection 
between the Ponzano-Regge ansatz and the quantization of the 
length in 3d loop gravity indicates immediately that in 4 
spacetime dimensions the naturally quantized geometrical 
quantities are not the lengths of the links, but rather areas and 
volumes of 2-cells and 3-cells of the triangulation 
\cite{RovelliPonzano,BarretWilliams}.  Therefore the natural 
coloring of the 4 dimensional state sum models should on the 
2-cells and 3-cells.  We will see in section \ref{foam} that 
this is precisely be the case.

\subsection{{\bf Old hopes $\rightarrow$ approximate 
theories}}

\subsubsection{Euclidean quantum gravity}

Euclidean quantum gravity is the approach based on a formal sum 
over Euclidean geometries
 \begin{equation}
 	Z \sim {\cal N} \int {\cal D}[g]\ e^{-\int d^{4}x \sqrt{g} 
 	R[g]}. 
 	\label{Hawk}
 \end{equation}
As far as I understand, Hawking and his close collaborators do 
not anymore view this approach as an attempt to directly define a 
fundamental theory.  The integral is badly ill defined, and does 
not lead to any known viable perturbation expansion.  However, 
the main ideas of this approach are still alive in several ways.

First, Hawking's picture of quantum gravity as a sum over 
spacetimes continues to provide a powerful intuitive reference 
point for most of the research related to quantum gravity.  
Indeed, many approaches can be sees as attempts to replace the 
ill defined and non-renormalizable formal integral (\ref{Hawk}) 
with a well defined expression.  The dynamical triangulation 
approach (Section \ref{discrete}) and the spin foam approach 
(Section \ref{foam}) are examples of  attempts to realize 
Hawking's intuition.  Influence of Euclidean quantum gravity can 
also be found in the Atiyah axioms for TQFT (Section \ref{tqft}).

Second, this approach can be used as an approximate method for 
describing certain regimes of nonperturbative quantum spacetime 
physics, even if the fundamental dynamics is given by a more 
complete theory.  In this spirit, Hawking and collaborators have 
continued the investigation of phenomena such as, for instance, 
pair creation of black holes in a background de Sitter spacetime.  
Hawking and Bousso, for example, have recently studied the 
evaporation and ``anti-evaporation'' of Schwarzschild-de Sitter 
black holes \cite{HawkingBousso}.

\subsubsection{Perturbative quantum gravity as effective theory, and 
the Woodard-Tsamis effect}

If expand classical GR around, say, the Minkowski metric, 
$g_{\mu\nu}(x)=\eta_{\mu\nu}+h_{\mu\nu}(x)$, and construct a 
conventional QFT for the field $h_{\mu\nu}(x)$, we obtain, as it 
is well know, a non renormalizable theory.  A small but 
intriguing group of papers has recently appeared, based on the 
proposal of treating this perturbative theory seriously, as a 
respectable low energy effective theory by its own.  This cannot 
solve the deep problem of understanding the world in general 
relativistic quantum terms.  But it can still be used for 
studying quantum properties of spacetime in some regimes.  This 
view has been advocated in a  convincing way by John 
Donoghue, who has developed effective field theory methods for 
extracting physics from non renormalizable quantum GR 
\cite{Donoghue}.

In this spirit, a particularly intriguing result is presented in 
a recent work by RP Woodard and NC Tsamis \cite{Woodard}.  
Woodard and Tsamis consider the gravitational back-reaction due 
to graviton's self energy on a cosmological background.  An 
explicit two-loop perturbative calculation shows that quantum 
gravitational effects act to slow the rate of expansion by an 
amount which becomes non-perturbatively large at late times.  The 
effect is infrared, and is not affected by the ultraviolet 
difficulties of the theory.  Besides being the only two loop 
calculation in quantum gravity (as far as I know) after the 
Sagnotti-Gorof non-renormalizability proof, this result is 
extremely interesting, because, if confirmed, it might represent 
an effect of quantum gravity with potentially observable 
consequences

\subsubsection{Quantum field theory on curved spacetime}

Unlike almost anything else described in this report, quantum 
field theory in curved spacetime is by now a reasonably 
established theory \cite{qftcs}, predicting physical phenomena of 
remarkable interest such as particle creation, vacuum 
polarization effects and Hawking's black-hole radiance 
\cite{hawkingrad}.  To be sure, there is no direct nor indirect 
experimental observation of any of these phenomena, but the 
theory is quite credible as an approximate theory, and many 
theorists in different fields would probably agree that these 
predicted phenomena are likely to be real. 

The most natural and general formulation of the theory is within 
the algebraic approach \cite{haag}, in which the primary objects 
are the local observables and the states of interest may all be 
treated on equal footing (as positive linear functionals on 
the algebra of local observables), even if they do not belong to 
the same Hilbert space.

In these last years there has been progress in the 
discussion of phenomena such as the instability of 
chronology horizons and on the issue of negative energies 
\cite{waldetal}.  Many problems are still open.  Interacting 
fields and renormalization are not yet completely 
understood, as far as I understand.  It is interesting to 
notice in this regard that the equivalence principle 
suggests that the problem of the ultraviolet divergences 
should be of the same nature as in flat space; so no 
obstruction for renormalization on curved spacetime is 
visible.  Nevertheless, the standard techniques for dealing 
with the problem are not viable, mostly for the 
impossibility of using Fourier decomposition, which is 
global in nature.  If ultraviolet divergences are a local 
phenomenon, why do we need global Fourier modes to deal with 
them?  A remarkable new development on these issues is the 
work of Brunetti and Fredenhagen \cite{brunetti}.  These 
authors have found a way to replace the requirement of the 
positive of energy (which is global) with a novel spectral 
principle based on the notion of wavefront (which is local).  
In this way, they make a substantial step towards the 
construction of a rigorous perturbation theory on curved 
spaces. The importance of a genuinely local formulation of 
QFT should probably not be underestimated.

The great merit of QFT on curved spacetime is that it has 
provided us with some very important lessons.  The key lesson is 
that in general one loses the notion of a single preferred 
quantum state that could be regarded as {\em the\/} ``vacuum''; 
and that the concept of ``particle'' becomes vague and/or 
observer-dependent in a gravitational context.  These conclusions 
are extremely solid, and I see no way of avoiding them.  In a 
gravitational context, {\em vacuum\/} and {\em particle\/} are 
necessarily ill defined or approximate concepts.  It is perhaps 
regrettable that this important lesson has not been yet absorbed 
by many scientists working in fundamental theoretical physics.

\subsection{{\bf ``Unorthodox'' approaches}}  

\subsubsection{Causal sets}

Raphael Sorkin vigorously advocates an approach to quantum 
gravity based on a sum over histories, where the histories are 
formed by discrete causal sets, or ``Posets'' \cite{sorkin}.  
Within this approach, he has discussed black hole entropy 
\cite{SorkinBH} and the cosmology constant problem. 
Sorkin's ideas have recently influenced various other 
directions. Markopoulou and Smolin have noticed that one 
naturally obtains a Poset structure in constructing a 
Lorentzian version of the spin foam models (see Section 
\ref{foam}).  Connections with noncommutative geometry have 
been explored in \cite{Landi2} by interpreting 
the partial ordering as a topology. 

\subsubsection{Finkelstein's ideas}

This year, David Finkelstein, original and radical thinker, has 
published his book, ``Quantum Relativity'', with the latest 
developments of his profound and fascinating re-thinking of the 
basis of quantum theory \cite{Finkelstein}.  The book contains a 
proposal on the possibility of connecting the elementary 
structure of spacetime with the internal variables (spin, color 
and isospin) of the elementary particles.  The suggestions has 
resonances with Alain Connes ideas (next Section).

\subsubsection{Twistors}

The twistor program has developed mostly on the classical and 
mathematical side.  Roger Penrose has presented intriguing and 
very promising steps ahead in this conference \cite{twistor}.  
Quantum gravity has been a major motivation for twistors; as far 
as I know, however, little development has happened on the 
quantum side of the program.

\section{New directions}

\subsection{{\bf Noncommutative geometry}} 
\label{noncomm} 

Noncommutative geometry is a research program in mathematics and 
physics which has recently received wide attention and raised 
much excitement.  The program is based on the idea that spacetime 
may have a noncommutative structure at the Planck scale.  A main 
driving force of this program is the radical, volcanic and 
extraordinary sequence of ideas of Alain Connes \cite{Connes}.

Connes' ideas are many, subtle and fascinating, and I cannot 
attempt to summarize them all here.  I mention only a few, 
particularly relevant for quantum gravity.  Connes observes that 
what we know about the structure of spacetime derives from our 
knowledge of the fundamental interactions: special relativity 
derives from a careful analysis of Maxwell theory; Newtonian 
spacetime and general relativity derived both from a careful 
analysis of the gravitational interaction.  Recently, we have 
learned to describe weak and strong interactions in terms of the 
$SU(3)\times SU(2)\times U(1)$ standard model.  Connes suggests 
that the standard model might hide information on the minute 
structure of spacetime as well.  By making the hypothesis that 
the standard model symmetries reflect the symmetry of a 
noncommutative microstructure of spacetime, Connes and Lott are 
able to construct an exceptionally simple and beautiful version 
of the standard model itself, with the impressive result that the 
Higgs field appears automatically, as the components of the Yang 
Mills connection in the internal ``noncommutative'' direction 
\cite{lott}.  The theory admits a natural extension in which the 
spacetime metric, or the gravitational field, is dynamical, 
leading to GR \cite{Chamseddine}.

What is a non-commutative spacetime?  The key idea is to use 
algebra instead of geometry in order to describe spaces.  
Consider a topological (Hausdorf) space $M$.  Consider all 
continuous functions $f$ on $M$.  These form an algebra $A$, 
because they can be multiplied and summed, and the algebra is 
commutative.  According to a celebrated result, due to Gel'fand, 
knowledge of the algebra $A$ is equivalent to knowledge of the 
space $M$, i.e.~$M$ can be reconstructed from $A$.  In 
particular, the points $x$ of the manifold can be obtained as the 
(one-dimensional) irreducible representations $x$ of $A$, which 
are all of the form $x(f)=f(x)$.  Thus, we can use the algebra of 
the functions, instead of using the space.  In a sense, notices 
Connes, the algebra is more physical, because we never deal with 
spacetime: we deal with fields, or coordinates, over spacetime.  
But one can capture Riemaniann geometry as well, algebraically.  
Consider the Hilbert space $H$ formed by all the spinor fields on 
a given Riemanian (spin) manifold.  Let $D$ be the (curved) Dirac 
operator, acting on $H$.  We can view $A$ as an algebra of 
(multiplicative) operators on $H$.  Now, from the triple 
$(H,A,D)$, which Connes calls ``spectral triple'', one can 
reconstruct the Riemaniann manifold.  In particular, it is not 
difficult to see that the distance between two points $x$ and $y$ 
can be obtained from these data by
\begin{equation}
	d(x,y)=sup_{\{f\in A, ||D,f||<1\}}\ |x(f)-y(f)|,
	\label{distance}
\end{equation}
a beautiful surprising algebraic definition of distance.  A  
non-commutative spacetime is the idea of describing spacetime by 
a spectral triple in which the algebra $A$ is a non-commutative 
algebra.

Remarkably, the gravitational field is captured, together with 
the Yang Mills field, and the Higgs fields, by a suitable Dirac 
operator $D$ \cite{Chamseddine}, and the full action is given 
simply by the trace of a very simple function of the Dirac 
operator.

Even if we disregard noncommutativity and the standard model, the 
above construction represents an intriguing re-formulation of 
conventional GR, in which the geometry is described by the Dirac 
operator instead than the metric tensor.  This formulation has 
been explored in \cite{Landi}, where it is noticed that the 
eigenvalues of the Dirac operator are diffeomorphism invariant 
functions of the geometry, and therefore represent true 
observables in (Euclidean) GR. Their Poisson bracket algebra can be 
explicitly computed in terms of the energy-momentum eigenspinors.  
Surprisingly, the Einstein equations turn out to be captured by 
the requirement that the energy momentum of the eigenspinors 
scale linearly with the eigenvalues.

Variants of Connes's version of the idea of non commutative 
geometry and noncommutative coordinates have been explored by 
many authors \cite{noncommutative} and intriguing connections 
with string theory have been suggested 
\cite{stringnonc,Froelich}.

A source of confusion about noncommutative geometry 
is the use of the expression ``quantum''.  In the 
mathematical parlance, one uses the expression ``quantization'' 
anytime one replaces a commutative structure with a 
noncommutative one, whether or not the non-commutativity has 
anything to do with quantum mechanics.\footnote{Noncommutativity 
can be completely unrelated to quantum theory, of course.  Boosts 
commute in Galilean relativity and do not commute in special 
relativity; but this does not mean that special relativity is by 
itself a quantum theory.} Models such as the Connes-Lott or the 
Chamseddine-Connes models are called ``quantum'' models by a 
mathematician, because they are based on a noncommutative 
algebra, but they are ``classical'' for a physicist, because 
they still need to be ``quantized'', in order to describe the physics 
of quantum mechanical phenomena.  If the model reduces to a 
standard Yang-Mills theory, then conventional QFT techniques can 
be used for the quantization.  Thus, for instance, the 
Connes-Lott models yields the conventional ``quantum'' standard 
model.  On the other hand, if the model includes a gravitational 
theory such as GR, which is non-renormalizable, then consistent 
quantization techniques are missing, and the difficulties of 
quantum GR are not solved, or mitigated, by just having a 
noncommutative manifold.  In such a model, the replacement of the 
commutative spacetime manifold with a noncommutative one is not 
sufficient to address the quantum physics of spacetime.

It is definitely too early to attempt a physical evaluation of the 
results obtained in this direction.  Many difficulties still 
separate the noncommutative approach from realistic physics.  The 
approach is inspired by Heisenberg intuition that physical 
observables become noncommutative at a deeper analysis, but so 
far a true merge with quantum theory is lacking.  Even in the 
classical regime, most of the research is so far in the unphysical 
Euclidean regime only. (What may replace equation (\ref{distance}) 
in the Lorentzian case?)  Nevertheless, this is a fresh set of 
new ideas, which should be taken very seriously, and which could 
lead to crucial advances.

\subsection{{\bf Null surface formulation}} 

A second new set of ideas comes from Kozameh, Newman and 
Frittelli \cite{nsf}.  These authors have discovered that the 
(conformal) information about the geometry is captured by 
suitable families of null hypersurfaces in spacetime, and have 
been able to reformulate GR as a theory of self interacting 
families of surfaces.  Since Carlos Kozameh has described this 
work in his plenary lecture in this conference \cite{carlos}, I 
will limit myself to one remark here.  A remarkable aspect 
of the theory is that physical information about the spacetime 
interior is transferred to null infinity, along null geodesics.  
Thus, the spacetime interior is described in terms of how we 
would (literally) ``see it'' from outside.  This description is 
diffeomorphism invariant, and addresses directly the relational 
localization characteristic of GR: the spacetime location of a 
region is determined dynamically by the gravitational field and 
is captured by when and where we see the spacetime region from 
infinity.  This idea may lead to  
interesting and physically relevant diffeomorphism invariant 
observables in quantum gravity.  A discussion of the quantum 
gravitational fuzziness of the spacetime points determined by 
this perspective can be found in \cite{fuzzy}.

\subsection{{\bf Spin foam models}}

\subsubsection{Topological quantum field theory} \label{tqft}

From the mathematical point of view, the problem of quantum 
gravity is to understand what is QFT on a differentiable manifold 
without metric (See section \ref{discussion}).  A class of well 
understood QFT's on manifolds exists.  These are the topological 
quantum field theories (TQFT).  Topological field theories are 
particularly simple field theories.  They have as many fields as 
gauges and therefore no local degree of freedom, but only a 
finite number of global degrees of freedom.  An example is GR in 
3 dimensions, say on a torus (the theory is equivalent to a Chern 
Simon theory).  In 3d, the Einstein equations require that the 
geometry is flat, so there are no gravitational waves.  
Nevertheless, a careful analysis reveals that the radii of the 
torus are dynamical variables, governed by the theory.  Witten 
has noticed that theories of this kind give rise to interesting 
quantum models \cite{wittentqft}, and \cite{atiyah} has provided 
a beautiful axiomatic definition of a TQFT. Concrete examples of 
TQFT have been constructed using hamiltonian, combinatorial and 
path integral methods.  The relevance of TQFT for quantum gravity 
has been suggested by many \cite{atiyah,tqft-qg} and the recent 
developments have confirmed these suggestions.

The expression ``TQFT'' is a bit ambiguous, and this fact has 
generated a certain confusion.  The TQFT's are diffeomorphism 
invariant QFT. Sometimes, the expression TQFT is used to indicate 
all diffeomorphism invariant QFT's.  This has lead to a 
widespread, but incorrect belief that any 
diffeomorphism invariant QFT has a finite number of degrees of 
freedom, unless the invariance is somehow broken, for instance 
dynamically.  This belief is wrong.  The problem of 
quantum gravity is precisely to define a diffeomorphism invariant 
QFT having an infinite number degrees of freedom and ``local'' 
excitations.  Locality in a gravity theory, however, is different 
from locality in conventional field theory.  Let me try to 
clarify this point, which is often source of confusion:
\begin{itemize} 
\item In a conventional 
field theory on a metric space, the degrees of freedom are 
local in the sense that they can be localized on the metric 
manifold (an electromagnetic wave is here or there in 
Minkowski space). 
\item  In a diffeomorphism invariant field 
theory such as general relativity, the degrees of freedom 
are still local (gravitational waves exist), but they are 
not localized with respect to the manifold.  They are 
nevertheless localized with respect to each other (a 
gravity wave is three meters apart from another gravity 
wave, or from a black hole).  
\item In a topological field theory, the degrees of 
freedom are not localized at all: they are global, and in 
finite number (the radius of a torus is not in a particular 
position on the torus).
\end{itemize}

Let me illustrate the main steps of the winding story.  The first 
TQFT directly related to quantum gravity was defined by Turaev 
and Viro \cite{TuraevViro}.  The Turaev-Viro model is a 
mathematically rigorous version of the 3d Ponzano-Regge quantum 
gravity model described in section \ref{pr}.  In the Turaev-Viro 
theory, the sum (\ref{PR}) is made finite by replacing $SU(2)$ 
with quantum $SU(2)_{q}$ (with a suitable $q$).  Since 
$SU(2)_{q}$ has a finite number if irreducible representations, 
this trick, suggested by Ooguri, makes the sum finite.  The 
extension of this model to four dimensions has been actively 
searched for a while and has finally been constructed by Louis 
Crane and David Yetter, again following Ooguri's ideas 
\cite{Ooguri,CraneYetter}.  The Crane-Yetter (CY) model is the 
first example of 4d TQFT. It is defined on a simplicial 
decomposition of the manifold.  The variables are spins 
(``colors'') attached to faces and tetrahedra of the 
simplicial complex.  Each 4-simplex contains 10 faces and 5 
tetrahedra, and therefore there are 15 spins associated to it.  
The action is defined in terms of the (quantum) Wigner 15-$j$ 
symbols, in the same manner in which the Ponzano-Regge action is 
constructed in terms of products of 6-$j$ symbols.  
\begin{equation}
Z \sim\!\! \sum_{coloring}\ \prod_{\mbox{\rm 4-}simplices}\!\!  
\mbox{\rm 15-}j({\rm color\ of\ the\  \mbox{\rm 4-}simplex}),
\label{CY}
\end{equation}
(I disregard various factors for simplicity).  
Crane and Yetter introduced their model independently 
from loop quantum gravity.  However, recall from Section \ref{pr} 
that loop quantum gravity suggests that in 4 dimensions the 
naturally discrete geometrical quantities are area and volume, 
and that it is natural to extend the Ponzano-Regge model to 4d by 
assigning colors to faces and tetrahedra. 

The CY model is not a quantization of 4d GR, nor could 
it be, being a TQFT in strict sense.  Rather, it can be 
formally derived  as a quantization of $SU(2)$ BF 
theory.  BF theory is a topological field theory with two fields, 
a connection $A$, with curvature $F$, and a two-form $B$ 
\cite{Horowitz}, with action
\begin{equation} 
               S[A,B] = \int B \wedge F. 
\end{equation}
However, there is a strict relation between GR and BF. If we add 
to $SO(3,1)$ BF theory the constraint that the two-form $B$ is 
the product of two tetrad one-forms
\begin{equation}
   	     B = E \wedge E, 
    \label{simple}
\end{equation}
we obtain precisely GR \cite{Plebanski,CapovillaDellJacobson}.  
This observation has lead many to suggest that a quantum theory 
of gravity could be constructed by a suitable modification of 
quantum BF theory \cite{BGGR}.  The suggestion has recently 
become very plausible, with the construction of the spin foam 
models, described below. 

\subsubsection{Spin foam models} \label{foam}

The key step was taken by Andrea Barbieri, studying the ``quantum 
geometry'' of the simplices that play a role in loop quantum 
gravity \cite{Barbieri}.  Barbieri discovered a simple relation 
between the quantum operators representing the areas of the faces 
of the tetrahedra.  This relation turns out to be the quantum 
version of the constraint (\ref{simple}), which turns BF theory 
into GR. Barret and Crane \cite{BarretCrane} added the 
Barbieri relation to (the $SO(3,1)$ version of) the CY model.  
This is equivalent to replacing the the 15-$j$ Wigner symbol, 
with a different function $A_{BC}$ of the colors of the 
4-simplex.  This replacement defines a ``modified TQFT'', which has a 
chance of having general relativity as its classical limit.  
Details and an introduction to the subject can be found in 
\cite{BaezSF}. 

The Barret-Crane model is not a TQFT in strict sense.  In 
particular, it is not independent from the triangulation.  
Thus, a continuum theory has to be formally defined by some 
suitable sum over triangulations
\begin{equation}
Z  \sim \!\!\!  \sum_{triang} \sum_{coloring} 
\prod_{\mbox{\rm 4-}simplices} \!\!\!\!
A_{BC}({\rm color\ of\ the\ \mbox{\rm 4-}simplex}).
	\label{BC}
\end{equation}
This essential aspect of the construction, however, 
is not yet understood. 

The striking fact is that the Barret Crane model can 
virtually be obtained also from loop quantum gravity.  This is an 
unexpected convergence of two very different lines of research.
Loop quantum gravity is formulated canonically in the frozen time 
formalism.  While the frozen time formalism is in principle 
complete, in practice it is cumbersome, and anti-intuitive.  Our 
intuition is four dimensional, not three dimensional.  An old 
problem in loop quantum gravity has been to derive a spacetime 
version of the theory.  A spacetime formulation of quantum 
mechanics is provided by the sum over histories.  A sum over 
histories can be derived from the hamiltonian formalism, as 
Feynman did originally.  Loop quantum gravity provides a 
mathematically well defined hamiltonian formalism, and one can 
therefore follow Feynman steps and construct a sum over histories 
quantum gravity starting from the loop formalism.  This has been 
done in \cite{rr}.  The sum over histories turns out to have the 
form of a sum over surfaces.

More precisely, the transition amplitude between two spin network 
states turns out to be given by a sum of terms, where each term 
can be represented by a (2d) branched ``colored'' surface in 
spacetime.  A branched colored surface is formed by elementary 
surface elements carrying a label, that meet on edges, also 
carrying a labeled; edges, in turn meet in vertices (or branching 
points). See Figure 3
\begin{figure} 
\centerline{\mbox{\epsfig{file=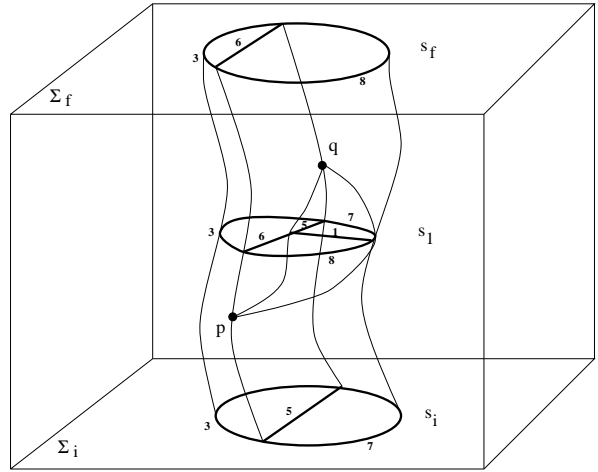}}} 
\caption{Figure 3: A branched surface with two vertices.}
\end{figure}
The contribution of one such surfaces to the sum over 
histories is the product of one term per each branching point of 
the surface.  The branching points represent the ``vertices'' of 
this theory, in the sense of Feynman. See Figure 4. 
\begin{figure} 
\centerline{\mbox{\epsfig{file=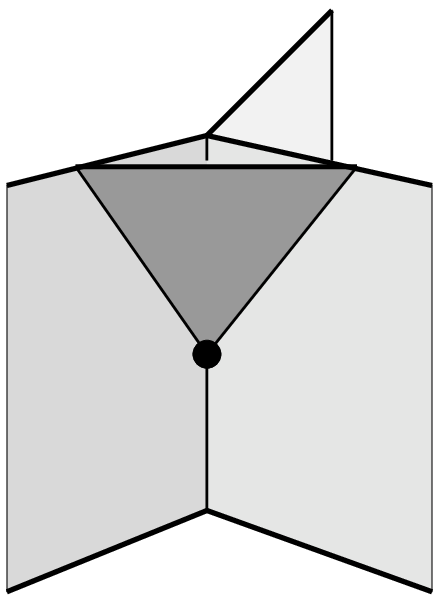}}} 
\caption{Figure 4: A simple vertex.}
\end{figure}
The contribution of each vertex can be computed algebraically 
from the ``colors'' (half integers) of the adjacent surface 
elements and edges.  Thus, spacetime loop quantum gravity is 
defined by the partition function
\begin{equation}
Z \sim \!\!\! \sum_{surfaces}\ \sum_{colorings}\ \prod_{vertices} 
\! A_{loop}({\rm color\ of\ the\ vertex})
	\label{Z}
\end{equation}
The vertex $A_{loop}$ is determined by a matrix elements of the 
hamiltonian constraint.  The fact that one obtains a sum over 
surfaces is not too surprising, since the time evolution of a 
loop is a surface.  Indeed, this was conjectured time ago by Baez 
and by Reisenberger.  The time evolution of a spin network (with 
colors on links and nodes) is a surface (with colors on surface 
elements and edges) and the hamiltonian constraint generates 
branching points in the same manner in which conventional 
hamiltonians generate the vertices of the Feynman diagrams.

What {\em is\/} surprising is that (\ref{Z}) has the same 
structure of the Barret Crane model (\ref{CY}).  To see this, 
simply notice that we can view each branched colored surface as 
located on the lattice {\em dual\/} to a triangulation (recall 
Figure 2).  Then each vertex correspond to a 4-simplex; the 
coloring of the two models matches exactly (elementary surfaces 
$\rightarrow$ faces, edges $\rightarrow$ tetrahedra); and summing 
over surfaces corresponds to summing over triangulations.  The 
main difference is the different weight at the vertices.  The 
Barret-Crane vertex $A_{BC}$ can be read as a covariant
definition a hamiltonian constraint in loop quantum 
gravity.

Thus, the spacetime formulation of loop quantum GR is a simple 
modification of a TQFT. This approach provides a 4d pictorial 
intuition of quantum spacetime, analogous to the Feynman graphs 
description of quantum field dynamics.  John Baez has introduced 
the term ``spin foam'' for the branched colored surfaces of the 
model, in honor of John Wheeler's intuitions on the quantum 
microstructure of spacetime.  Spin foams are a precise 
mathematical implementation of Wheeler's ``spacetime foam'' 
suggestions. Markopoulou and Smolin have explored 
the Lorentzian version of the spin foam models \cite{Markopoulou}.  

This direction is very recent.  It is certainly far too early to 
attempt an evaluation.  Many aspects of these models are still 
obscure.  But the spacetime foam models may turn out among the 
most promising recent development in quantum gravity.

\vskip1cm

This concludes the survey of the main approaches to quantum 
gravity. 

\section{Black hole entropy} \label{BH}

A focal point of the research in quantum gravity in the last 
years has been the discussion of black hole (BH) entropy.  This 
problem has been discussed from a large variety of perspectives 
and within many different research programs.  

Let me very briefly recall the origin of the problem.  In {\em 
classical\/} GR, future event horizons behave in a manner that 
has a peculiar thermodynamical flavor.  This remark, together 
with a detailed physical analysis of the behavior of hot matter 
in the vicinity of horizons, prompted Bekenstein, over 20 years 
ago, to suggest that there is entropy associated to every 
horizon.  The suggestion was first consider ridicule, because it 
implies that a black hole is hot and radiates.  But then Steven 
Hawking, in a celebrated work \cite{hawkingrad}, showed that QFT 
in curved spacetime predicts that a black hole emits thermal 
radiation, precisely at the temperature predicted by Bekenstein, 
and Bekenstein courageous suggestion was fully vindicated.  Since 
then, the entropy of a BH has been indirectly computed in a 
surprising variety of manners, to the point that BH entropy and 
BH radiance are now considered almost an established fact by the 
community, although, of course, they were never observed nor, 
presumably, they are going to be observed soon.  This confidence, 
perhaps a bit surprising to outsiders, is related to the fact 
thermodynamics is powerful in indicating general properties of 
systems, even if we do not control its microphysics.  Many hope 
that the Bekenstein-Hawking radiation could play for quantum 
gravity a role analogous to the role played by the black body 
radiation for quantum mechanics.

Thus, indirect arguments indicate that a Schwarzschild BH 
has an entropy 
\begin{equation}
	S=\frac{1}{4}\ \frac{A}{\hbar G}
	\label{entropy}
\end{equation}
The challenge is to derive this formula from first principles.  A 
surprisingly large number of derivations of this formula have 
appeared in the last years.

\begin{description}
\item [String theory.]  In string theory, one can count the 
number of string states that have the same mass and the same 
charges at infinity as an extremal BH. (An extremal BH is a BH 
with as much charge as possible -- astrophysically, an 
extremal BH is a highly improbable object).  Since a BH is a 
nonperturbative object, the calculation refers to the 
nonperturbative regime, where string theory is poorly understood.  
But it can nevertheless be completed, thanks to a trick.
At fixed mass and charge at infinity, if the coupling 
constant is large, there is a BH, but we do not control the 
theory; if the coupling is weak there is no BH, but the theory is 
in the perturbative, and we can count states with given mass and 
charges.  Thanks to a 
(super-) symmetry of string theory, quantum states corresponding 
to extremal BH (BPS states) have the property that their number 
does not depend on the strength of the coupling constant.  
Therefore we can count them in the limit of weak coupling, and 
be confident that the counting holds at strong coupling as well.  
In this way we can compute how many states in the theory (at strong 
coupling) correspond to a black hole geometry 
\cite{stringentropy,mal}.  The striking results is that if we 
interpret BH entropy as generated by the number of such states 
($S = k \ln N$) we obtain the correct Bekenstein Hawking formula, 
with the correct 1/4 factor. 

The derivation has been extended outside the 
extremal case \cite{HorowitzStrominger}, but I am not 
aware, so far, of a  result for non extremal BH's 
as clean and compelling as the result for the extremal case.
The Hawking radiation rate itself can be derived from the 
string picture (for the near extremal black holes) 
\cite{DasMathur}. For long wavelength radiation, one can also  
calculate the `grey body factor', namely corrections to the 
thermal spectrum due to frequency-dependent potential 
barriers outside the horizon, which filter the initially 
blackbody spectrum emanating from the horizon \cite{Maldacena}.

The derivation is striking, but it leaves many open questions.  
Are the distinct BH states that enter the counting 
distinguishable form each other, for an observer at infinity?  If 
they are not distinguishable, how can they give rise to 
thermodynamical behavior?  (Entropy is the number of {\em 
distinguishable\/} states.)  If yes, which observable can 
distinguish them?  Is a black hole in string theory not really 
black?  (See \cite{dongary}.)  More in general, why is there this 
consistency between string theory, QFT in curved space and 
classical GR? How does thermodynamics, horizons and quantum 
theory interplay? 

Of course, one cannot view the BH entropy derivation as 
experimental support for string theory: BH radiance has never 
been observed, and, even if it is observed, BH radiance was 
predicted by Hawking, not by string theory, and it is just a 
consequence of QFT on a curved background plus classical GR, not 
of string theory.  Any theory of quantum gravity consistent with 
the QFT in curved spacetime limit should yield the BH entropy.  
What the string derivation does show is that string theory is 
indeed consist with GR and with QFT on curved space, even in the 
strong field regime, where the theory is poorly understood, and 
at least as far as the extremal case in concerned.  The 
derivation of the entropy formula for the extremal BH represents 
a definite success of string theory. 

\item[Surface states.]  Fifteen years ago, York suggested that 
the degrees of freedom associated to BH entropy could be 
interpreted as fluctuations of the position of the event horizon 
\cite{York}.  Thus, they could reside on the horizon itself.  
This suggestion has recently become precise.  The new idea is 
that the horizon (in a precise sense) breaks diffeomorphism 
invariance locally, and this fact generates quantum states on the 
BH surface \cite{Carlip,bala,Teitel}.  These states are called 
edge states, or surface states, and can naturally be described in 
terms of a topological theory on the horizon.  Balachandran, 
Chandar and Momen have derived the existence of surfaces states 
in 3+1 gravity, and showed that these are described by a surface 
TQFT \cite{bala}.  Steven Carlip has shown that this ideas leads 
to a computation of the BH entropy in 2+1 gravity, obtaining the 
correct 1/4 factor \cite{Carlip}.  The surface states idea is 
also at the root of the loop quantum gravity derivations 
described below.

\item [Loop Quantum Gravity.]  Kirill Krasnov has introduced 
statistical techniques for counting loop microstates 
\cite{loopentropy1} and has opened the study of BH entropy within 
loop quantum gravity.  There are two derivations of BH entropy 
from loop quantum gravity.  One \cite{loopentropy2,loopentropy1} 
is based on a semiclassical analysis of the physics of a hot 
black hole.  This analysis suggests that the area of the horizon 
does not change while it is thermally ``shaking''.  This implies 
that the thermal properties of the BH are governed by the number 
of microstates of the horizon having the same area.  The 
apparatus of loop quantum gravity is then employed to compute 
this number, which turns out to be finite, because of the Planck 
scale discreteness implied by the existence of the quanta of 
geometry (Section \ref{quanta}).  The number of relevant states 
is essentially obtained from the number of the eigenvalues in 
equation (\ref{area}) hat have a given area.  In the second 
approach \cite{loopentropy3}, one analyzes the classical theory 
{\em outside\/} the horizon treating the horizon as a boundary.  
A suitable quantization of this theory yields surface states, 
which turn out to be counted by an effective Chern Simon theory 
on the boundary, thus recovering the ideas of Balachandran and 
collaborators.  In both derivations, one obtains, using 
Eq.(\ref{area}), that the entropy is proportional to the area in 
Planck units.  However, loop quantum gravity does not fix the 
(finite) constant of proportionality, because of the parameter 
$\gamma$ in (\ref{area}), a finite free dimensionless parameter 
not determined by the theory, first noticed by Immirzi 
\cite{Immirzi,RT}.

In comparison with the string derivation, the loop derivation is 
weaker because it does not determines univocally the 1/4 factor 
of Eq.(\ref{entropy}) and it stronger because it works naturally 
for ``realistic'' BH's, such as Schwarzschild.

\item[Entanglement entropy.]
An old idea about BH entropy, first considered by Bombelli, 
Koul, J Lee and Sorkin is that it is the effect of the short 
scale quantum entanglement between the two sides of the 
horizon \cite{SorkinBombelli}.  A similar idea was 
independently proposed by Frolov and Novikov, who suggest 
that BH entropy reflects the degeneracy with respect to 
different quantum states which exist inside the black hole, 
where inside modes contribute only if they are correlated 
with external modes \cite{Frolov1}.  The idea of 
entanglement entropy has been recently analyzed in detail in 
\cite{kodama}, where it is suggested that, suitably 
interpreted, the idea might still be valid.

\item[Induced gravity.] Frolov and Fursaev have developed the 
idea of entanglement entropy by applying it within 
Sakharov's induced gravity theories, following a suggestion 
by Ted Jacobson.  The idea of using induced gravity theories 
is motivated by the fact that the bare gravitational 
constant gets renormalized in the computation of the 
entanglement entropy, yielding a divergent entropy 
\cite{Frolov2}.  In induced gravity theories, there is no 
bare gravitational constant, and one may obtain the correct 
finite answer.  The idea is that a (still unknown) 
fundamental theory which induces the correct low energy 
gravity should allow a representation in terms of an 
infinite set of fields which could play the role of the 
induced gravity  constituents.

\item[Bekenstein's model.]  Bekenstein, the ``inventor'' of BH 
entropy, has recently analyzed the quantum structure of BH's, 
using the idea that a BH can be treated as simple quantum 
objects.  This quantum object has, presumably, quantized energy 
levels, or horizon-area levels, since energy and area are related 
for a BH. Therefore it emits energy in quantum jumps, as atoms 
do.  An interesting consequence of this approach 
\cite{BekensteinMukhanov} is that if the horizon area levels are 
equispaced, the emitted radiation differs strongly from the 
thermal one predicted by Hawking, and presents macroscopically 
spaced spectral lines.  This would be extremely interesting, 
because these spectral lines might represent observable 
macroscopic QG effects.  Unfortunately, several more complete 
approaches, and in particular, loop quantum gravity, do not lead 
to equispaced area levels (see Eq.(\ref{area})).  With a more 
complicated area spectrum the emitted radiation is effectively 
thermal \cite{marcelo,abhayarea}, as predicted by Hawking.  
Nevertheless, the Bekenstein-Mukhanov effect remains an 
intriguing idea: for instance, it has been suggested that 
it might actually resurrect in loop quantum gravity for dynamical 
reasons.

\item['t Hooft's] {\bf ``S-matrix ansatz'' and ``holographic 
principle''.}  In conjunction with his discussion of BH's 
radiation, Steven Hawking has long claimed that BH's violate 
ordinary quantum mechanics, in the sense that a pure state 
can evolve into a mixed state in the presence of a BH. More 
precisely, the evolution from $t=-\infty$ to $t=+\infty$ is 
not given by the $S$-matrix acting on physical states, but 
rather by an operator, which he calls \$-matrix, acting on 
density matrices.  Gerard 't Hooft has been disputing this 
view for a long time, maintaining that the evolution should 
still be given by an $S$-matrix.  't Hooft observes that if 
one assumes the validity of conventional quantum field 
theory in the vicinity of the horizon, one does not find a 
quantum mechanical description of the BH that resembles that 
of conventional forms of matter.  Instead, he considers the 
alternative assumption that a BH can be described as an 
ordinary object within unitary quantum theory.  The 
assumption of the existence of an ordinary $S$-matrix has far 
reaching consequences on the nature of space-time, and even 
on the description of the degrees of freedom in ordinary 
flat space-time.  In particular, the fact that all 
microstates are located on the horizon implies a puzzling 
property of space-time itself, denoted the holographic 
principle.  According to this principle, the combination of 
quantum mechanics and gravity requires the three dimensional 
world to be an image of data that can be stored on a two 
dimensional projection -- much like a holographic image.  The 
two dimensional description only requires one discrete 
degree of freedom per Planck area and yet it is rich enough 
to describe all three dimensional phenomena.  These views 
have recently been summarized in Ref.~\cite{tHooft}.  Suskind 
has explored some consequences of `t Hooft's holographic 
principle, showing that it implies that particles must grow 
in size as their momenta increase far above the Planck 
scale, a phenomenon previously discussed in the context of 
string theory, thus opening a possible connection between 't
Hooft views and string theory \cite{Susskind}.

\item[Trans-Planckian frequencies.]  
Work by Unruh and Jacobson has provided interesting insight into 
how the prediction of Hawking radiation apparently is not 
affected by modifications of the theory at ultra-high 
frequencies.  If the modes of the Hawking radiance are red 
shifted emerging from a BH, one might imagine that finite 
frequencies at infinity derive from arbitrary high frequencies at 
the horizon.  But if spacetime is discrete, arbitrary high modes 
do not exist.  One can get out from this apparent paradox by 
observing that the outgoing modes do not arise from high 
frequency modes at the horizon, but from ingoing modes, through a 
process of ``mode conversion'' which is well known in plasma 
physics and in condensed matter physics \cite{JacobsonBH}.
    
\item[Others.]  Several others results on black hole entropy 
exist \cite{SorkinBH}.  I do not have the 
space or the competence for an exhaustive list.  For a 
recent overview, see Ted Jacobson review of the BH entropy 
section of the MG8 \cite{tedMG8}.

\end{description}

The above list shows that there is a rather large number of 
research programs on BH entropy.  Many of these programs 
claim that a key for solving the puzzle has been found.  
However most research program ignore the others.  
Presumably, what is required now is a detailed comparison of 
the various ideas.

\section{The problem of quantum gravity. A discussion} 
\label{discussion}

\setcounter{subsubsection}{0}

The problem that the research on quantum gravity addresses is 
simply formulated: finding a fundamental theoretical description 
of the physics of the gravitational field in the regime in which 
its quantum mechanical properties cannot be disregarded.

This problem, however, is interpreted in surprisingly different 
manners by physicists with different cultural backgrounds.  There are 
two main interpretations of this problem, driving the present 
research: the particle physicist's one and the relativist's one.

\subsubsection{The problem, as seen by a high energy physicist}

High energy physics has obtained spectacular successes 
during this century, culminated with the establishment of 
quantum field theory and of the $SU(3)\times SU(2)\times 
U(1)$ standard model.  The standard model encompasses 
virtually everything we can physically measure -- except 
gravitational phenomena.  From the point of view of a 
particle physicist, gravity is simply the last and weakest 
of the interactions.  It is natural to try to understand its 
quantum properties using the same strategy that has been so 
successful for the rest of microphysics, or variants of this 
strategy.  The search for a conventional quantum field 
theory capable of embracing gravity has spanned several 
decades and, through a curious sequence of twists, 
excitements and disappointments, has lead to string theory.

For a physicist with a high energy background, the problem 
of quantum gravity is thus reduced to an aspect of the 
problem of understanding what is the mysterious 
nonperturbative theory that has perturbative string theory 
as its perturbation expansion, and how to extract 
information on Planck scale physics from it.

In string theory, gravity is just one of the excitations of a 
string (or other extended object) living over some metric space.  
The existence of such background metric space, over which the 
theory is defined, is needed for the formulation of the theory, 
not just in perturbative string theory, but also in most of the 
recent attempts of a non-perturbative definition of the theory, 
as I argued in section \ref{string}.

\subsubsection{The problem, as seen by a relativist} 

For a relativist the idea of a fundamental 
description of gravity in terms of physical excitations over 
a metric space sounds incorrect.  The key lesson of GR is that 
there is no background metric over which physics happens 
(except in approximations).  The gravitational 
field is the same physical object as the spacetime itself, 
and therefore quantum gravity is the theory of the quantum 
microstructure of spacetime.  To understand quantum gravity 
we have to understand what is quantum spacetime.

More precisely, for a relativist, GR is much more than the 
field theory of a particular force.  Rather, it is the 
discovery that certain classical notions about space and 
time are not adequate at the fundamental level; and require 
a deep modifications.  One of such inadequate notions is 
precisely the notion of a background metric space (flat or 
curved), {\em over\/} which physics happens.  It is this 
conceptual shift that has led to the understanding of 
relativistic gravity, to the discovery of black holes, to 
relativistic astrophysics and to modern cosmology.  For a 
relativist, quantum gravity is the problem of merging {\em 
this\/} conceptual novelty with quantum field theory.

From Newton to the beginning of this century, physics has 
been founded over a small number of key notions such as 
space, time, causality and matter.  In the first quarter of 
this century, quantum theory and general relativity have 
modified this foundation in depth.  The two theories have 
obtained solid success and vast experimental corroboration, 
and can be now considered as well established knowledge.  
Each of the two theories modifies the conceptual foundation 
of classical physics in a (more or less) internally 
consistent manner.  However, we do not have a novel 
conceptualization of the physical world capable of embracing 
both theories.  For a relativist, the challenge of quantum 
gravity is the problem of bringing this vast conceptual 
revolution, started with quantum mechanics and with general 
relativity, to a conclusion and to a new synthesis.

Unlike perturbative or nonperturbative string theory, 
relativist's quantum gravity theories tend to be formulated 
without a background spacetime, and are direct attempts to grasp 
what is quantum spacetime at the fundamental level.

\subsubsection{What is quantum spacetime?}\label{rel}

General relativity has taught us that the spacetime metric 
is dynamical, like the rest of the physical entities.  From 
quantum mechanics we have learned that all dynamical entities 
have quantum properties (undergo quantum fluctuations, are 
quantized, namely they tend to manifest themselves in small 
quanta at short scale, and so on).  These quantum properties 
are captured by the basic formalism of quantum mechanics, in 
its various versions.  Thus, we expect spacetime metric to 
be subject to Heisenberg's uncertainty principle, to come in 
small packet, or quanta of spacetime, and so on.  Spacetime 
metric should then only exist as an expectation value of 
some quantum variable.

But we have learned another more general lesson 
from GR: that spacetime location is relational only.  This 
is a distinct idea from the fact that the metric is 
dynamical.  Mathematically, this physical idea is captured 
by the active active diff invariance of the Einstein 
equations.  (Einstein searched for {\em non-diff 
invariant\/} equations for a dynamical metric and for the Riemann 
tensor from 1912 to 1915, before understanding the need of 
active diff invariance in the theory.)  Active diff 
invariance means that the theory is invariant under a 
diffeomorphism on the dynamical fields of the theory 
(not on every object of the theory: any theory, suitably 
formulated is trivially invariant under a diffeomorphism on 
all its objects).  Physically, diff invariance 
has a profound and far reaching meaning. This meaning is subtle, 
and even today, 75 years after the discovery of GR, it is 
sometimes missed by theoretical physicists, particularly 
physicists without a GR background.

A non diff-invariant theory of a system $S$ describes the 
evolution of the objects in $S$ with respect to a reference 
system made by objects external to $S$.  A diff-invariant theory 
of a system $S$ describes the dynamics of the objects in $S$ with 
respect to each others.  In particular, localization is defined 
only internally, relationally.  Objects are somewhere only with 
respects to other dynamical objects of the theory, not with 
respect to an external reference system.  The electromagnetic 
field of Maxwell theory is located somewhere in spacetime.  The 
gravitational field is not located  in spacetime: it is 
with respect to {\em it\/} that things are localized.
To put it pictorially, pre-GR physics describes the motion of 
physical entities over the stage formed by a non-dynamical 
spacetime.  While general relativistic physics describes the 
dynamics of the stage itself.  The stage does not ``move'' 
over a background. It ``moves'' with respects to itself. 
Therefore, what we need in quantum gravity is {\em a relational 
notion of a quantum spacetime}.

General quantum theory does not seem to contain any element 
incompatible with this physical picture.  On the other hand, 
conventional quantum {\em field\/} theory does, because it is 
formulated as a theory of the motion of small excitations over a 
background.  Thus, to merge general relativity and quantum 
mechanics we need a quantum theory for a field system, but 
different from conventional QFT over a given metric space.  
General relativity, as a classical field theory, is not defined 
over a metric space, but over a space with a much weaker 
structure: a differentiable manifold.  Similarly, in quantum 
gravity we presumably need a QFT that lives over a manifold.  
Mathematically, the challenge of quantum gravity can therefore be 
seen as the challenge of understanding how to consistently define 
a QFT over a manifold, as opposite to a QFT over a metric space.  
The theory must respects the manifold invariance, namely active 
diffeomorphism.  This means that the location of states on the 
manifold is irrelevant.  

This idea was beautifully expressed by Roger Penrose in the work 
in which he introduced spin networks \cite{Penrose}.

\begin{quotation}
 ``A reformulation is suggested in which quantities normally 
  requiring continuous coordinates for their description are 
  eliminated from primary consideration.  In particular, 
  since space and time have therefore to be eliminated, what 
  might be called a form of Mach's principle be invoked: a 
  relationship of an object to some background space should 
  not be considered -- only relationships of objects to 
  each other can have significance.''	
\end{quotation}

Several of the research programs described above realize this 
program to a smaller or larger extent.  In particular, recall 
that the spin network states of loop quantum gravity (See Figure 
1) are not excitations {\em over\/} spacetime.  They are 
excitations {\em of\/} spacetime.  This relational aspect of 
quantum gravitational states is one of the most intriguing 
aspects emerging from the theory. 

\subsubsection{Quantum spacetime, other aspects}

The old idea of a lower bound of the divisibility of space around 
the Planck scale has been strongly reinforced in the last years.  
Loop quantum gravity has provided {\em quantitative\/} evidence 
in this sense, thanks to the computation of the quanta of area 
and volume.  The same idea appears in string theory, in certain 
aspects of non commutative geometry, in Sorkin's poset theory, 
and in other approaches \cite{Landi2}.

Notice, in this regard, that spacetime is discrete in the quantum 
sense.  It is not ``made by discrete quanta'', in the sense in 
which an electron is not ``made by Bohr orbitals''.  A generic 
spacetime is a quantum superposition of discretized states.  
Outcome of measurements can be discrete, expectation values are 
continuous.

Physically, one can view the Planck scale discreteness as 
produced by short scale quantum fluctuations: at the scale at 
which these are sufficiently strong, the virtual energy density 
is sufficient to produce micro black holes.  In other words, flat 
spacetime is unstable at short scale.  A recent variational 
computation \cite{Preparata} confirms this idea by showing that 
flat spacetime has higher energy than a spacetime made of Planck 
scale black holes. 

Another old idea that has consequently been reinforced in the 
last years is that the perturbative picture of a Minkowski space 
with real and virtual gravitons is not appropriate at the Planck 
scale.  For instance, the string black hole computation does not 
works because the weak coupling expansion reaches the relevant 
regime, but because there is a special case in which one can 
independently argue that the number of states is the same at weak 
and strong coupling.

In general, thus, there seem to be a certain convergence in the 
emerging physical picture of Planck scale quantum spacetime.  
However, we are far from a point in which we can say that we 
understand the structure of quantum spacetime, and many general 
questions remain open.  In loop quantum gravity, a credible state 
representing Minkowski has not been found yet.  In string theory, 
there are too many vacua and the theory does not seem to have 
much predictivity about the details of the Planck scale structure 
of spacetime.  In the discrete approach, it is not yet clear 
whether the phase transition gives rise to a large scale theory, 
and, if so, whether the discrete structure of the triangulations 
leaves a physical remnant (in QCD it does not, of course).  A 
general problem is the precise relation between spacetime's 
microphysics and macrophysics.  Do we expect a full fledged 
renormalization group to play a role?  Do we expect large scale 
physics to be insensitive to the details of the microphysics, as 
happens in renormalizable QFT? Or the existence of a physical 
cutoff kills this idea?  Smolin, has suggested that the existence 
of a phase transition should not be a defining property of the 
theory, but rather a property of certain states in the theory, 
the ones that yield macroscopic spacetimes instead of Planck 
scale clots.  Much is still unclear about quantum spacetime.

\section{Relation between quantum gravity and other major open 
problems in fundamental physics}

When one contemplates two deep problems, one is immediately 
tempted to speculate that they are related.  Quantum gravity has 
been asked, at some time or the other, to take charge of almost 
every other open problem in theoretical physics (and 
beyond).  Here is a list of problems that at some time or another 
have been connected to quantum gravity.

It is important to remark that, with few important exceptions, 
these problems might very well turn out to be unrelated to 
quantum gravity.  The history of physics is full of examples of 
two problems solved together (say: understanding the nature of 
light and uniting electricity to magnetism).  But it is also 
full of disappointed great hopes of getting two results with one 
stroke (say: finding a theory of the strong interactions and 
getting rid of ultraviolet divergences and infinite 
renormalization).  In particular, the fact that a proposed 
solution to the quantum gravity puzzle does not address this or 
that of the following problems is definitely not an indication 
it is physically wrong.  QCD was initially criticized as a 
theory of strong interactions because it did not solve the 
puzzles raised by renormalization theory.  We should not 
repeat that mistake.

I begin with various issues related to quantum mechanics, which
are sometimes confused with each other.  

\begin{description}

\item[Quantum Cosmology.]
There is widespread confusion between quantum cosmology and 
quantum gravity.  Quantum cosmology is the theory of the entire 
universe as a quantum system without external observer 
\cite{HartleQC,Hartle}.  The problem of quantum cosmology exists 
with or without gravity.  Quantum gravity is the theory of one 
dynamical entity: the quantum gravitational field (or the 
spacetime metric): just one entity among the many.  We can assume 
that we have a classical observer with a classical measuring 
apparatus measuring quantum gravitational phenomena, and 
therefore we can formulate quantum gravity disregarding quantum 
cosmology.  In particular, the physics of a Planck size small 
cube is governed by quantum gravity and, presumably, has no 
cosmological implications.  Quantum cosmology addresses an 
extremely general and important open question.  But that question 
is not necessarily tied to quantum gravity.

\item[Quantum theory ``without time''. Unitarity.]
The relational character of GR described in Section \ref{rel} is 
reflected in the peculiar role of time in gravity.  GR does not 
describe evolution with respect to an external time, but only 
relative evolution of physical variables with respect to each 
other.  In other words, temporal localization is relational like 
spatial localization.  This is reflected in the fact that the 
theory has no hamiltonian (unless particular structures are 
added), but only a ``hamiltonian'' constraint.  Conventional 
quantum mechanics needs to be adapted to this way of treating 
time.  There are several ways of doing so.  Sum over histories 
may be a particularly suitable way of formulating such 
``generalized'' quantum mechanics in a gravitational context, as 
suggested by the work of Jim Hartle \cite{Hartle}; canonical 
methods are viable as well \cite{RovelliTime}.  For an extensive 
discussion of the problem and its many subtleties, see 
\cite{IshamTime}.

Opinions diverge on whether a definition of time evolution must 
be unitary in nonperturbative quantum gravity.  If we assume 
asymptotic flatness, then there is a preferred time at infinity 
and Poincare' symmetry at infinity implies unitarity.  Outside 
this case, the issue is much more delicate.  Unitarity is needed 
for the consistency of a theory in flat space.  But the 
requirement of unitarity should probably not be mistaken for a 
general consistency requirement, and erroneously extended from 
the flat space domain, where there is an external time, to the 
quantum gravity domain, where there is no external time.  In GR, 
one can describe evolution with respect to a rather arbitrarily 
chosen physical time variable $T$.  There is no reason for a 
$T$-dependent operator $A(T)$ to be unitarily related to $A(0)$.  
Lack of unitarity simply means that the time evolution of a 
complete set of commuting observables may fail to be a complete 
set of commuting observables.  This is an obstruction for the 
definition of a Shr\"odinger picture of time evolution, but the 
Heisenberg picture \cite{RovelliTime}, or the path integral 
formulation \cite{Hartle}, may nevertheless be consistent.

\item[Structure ] {\bf and interpretation of quantum mechanics.  
Topos theory.} It has been often suggested that the much debated 
interpretative difficulties of quantum theory may be related to 
quantum gravity, or that the very structure of quantum mechanics 
might have to undergo a substantial revision in order to include 
GR. In Ted Newman's views, for instance, the gravitational field 
is so physically different from any other field, that 
conventional quantizations methods,
\begin{quote}
	``another form of orthodoxy'', 
\end{quote}
as Ted calls them, are unlikely to succeed.  Thus, Newman 
advocates the need of a substantial revision of quantum theory in 
order to understand quantum gravity, and expects that the 
mysteries of quantum gravity and the mysteries of quantum 
mechanics be intertwined.

Certainly, quantum gravity and quantum cosmology have played an 
indirect role in the effort to understand quantum theory.  If 
quantum theory has to play the role of general theory of 
mechanics, it certainly has to be general enough to encompass the 
peculiar features of gravitational theories as well.  In 
particular, the consistent-histories approach to quantum theory 
\cite{consistent} was motivated in part by the search for an 
interpretation viable in a context where the microstructure of 
spacetime itself is subject of quantum effects.

A fascinating development in this direction is the recent work of 
Chris Isham on the relevance of topos-theory in the histories 
formulation of quantum mechanics \cite{Topos}.  The main idea is 
to assign to each proposition $P$ a truth value defined as the 
set (the ``sieve'') of all consistent families of histories 
within which $P$ holds.  The set of all such sieves forms a 
logical algebra, albeit one that contains more than just the 
values `true' and `false'.  This algebra is naturally described 
by topos theory.  Isham's topos-theoretical formulation of 
quantum mechanics is motivated in part by the desire of extending 
quantum theory to contexts in which a classical spacetime does 
not exist.  More in general, topos theory has a strongly 
relational flavor and emphasizes relational aspects of quantum 
theory.  (Relational aspects of quantum theory are discusses also 
in \cite{relational}.)  The existence of a connection between 
such relational aspects of quantum theory and relational aspects 
of GR (Section \ref{rel}) has been explicitly suggested in 
Refs.~\cite{half,crane}, and might represent a window over a 
still unexplored realm.  These difficult issues are still very 
poorly understood, but they could turn out to be crucial for 
future developments. 

\item[Wave function collapse.]
A direct implementation of the idea that the mysteries of quantum 
gravity and the mysteries of quantum mechanics can be related is 
Penrose's suggestion that the wave function collapse may be a 
gravitational phenomenon.  Penrose's idea is that there may be a 
nonlinear dynamical mechanism that forbids quantum superpositions 
of (``too different'') spacetimes.  A fact that perhaps supports 
the speculation is the disconcerting value of the Planck mass.  The 
Planck mass, 22 micrograms, lies approximately at the boundary 
between the light objects that we see behaving mostly quantum 
mechanically and the heavy objects that we see behaving mostly 
classically.  Since the Planck mass contains the Newton constant, 
this coincidence might be read as an indication that gravity 
plays a role in a hypothetical transition between quantum and 
classical physics.  Consider an extended body with mass $M$ in a 
quantum superposition of two states $\Psi_{1}$ and $\Psi_{2}$ in 
which the center of mass is, respectively, in the positions 
$X_{1}$ and $X_{2}$.  Let $U_{grav}$ be the gravitational 
potential energy that two distinct such bodies would have if they 
were in $X_{1}$ and $X_{2}$.  Penrose suggests that the quantum 
superposition $\Psi_{1} + \Psi_{2}$ is unstable and naturally 
decays through some not yet known dynamics to either $\Psi_{1}$ 
or $\Psi_{2}$, with a decay time
\begin{equation}
	t_{collapse} \sim \frac{\hbar}{U_{grav}}. 
	\label{penrose}
\end{equation}
The decay time (\ref{penrose}) turns out to be surprisingly 
realistic, as one can easily compute: a proton can be in a 
quantum superposition for eons, a drop of water decays 
extremely fast, and the transition region in which the decay 
time is of the order of seconds is precisely in the regime in 
which we encounter the boundary between classical and quantum 
behavior. 

The most interesting aspect of Penrose's idea is that it can be 
tested in principle, and perhaps even in practice.  Antony 
Zeilinger has announced in his plenary talk in this conference 
\cite{zeilinger} that he will try to test this prediction in the 
laboratory.  Most physicists would probably expect that 
conventional quantum mechanics will once more turn out to be 
exactly followed by nature, and the formula (\ref{penrose}) will 
be disproved. But it is certainly worthwhile checking. 

\item[Unifications of all interactions] {\bf and ``Theory of 
Everything''.} String theory represents a tentative solution of 
the quantum gravity problem, but also of the problem of unifying 
all presently known fundamental physics.  This is a fascinating 
and attractive aspect of string theory.  On the other hand, this 
is not a reason for discarding alternatives.  The idea that 
quantum gravity can be understood only in conjunctions with other 
matter fields is an interesting hypothesis, not an established 
truth.

\item[Origin of the Universe.]  It is likely that a sound quantum 
theory of gravity will be needed to understand the physics 
of the Big Bang.  The converse is probably not true: we 
should be able to understand the small scale structure of 
spacetime even if we do not yet understand  the origin of the 
Universe.

\item[Ultraviolet divergences.]  As already mentioned, a great 
hope during the search for the fundamental theory of the strong 
interactions was to get rid of the QFT's ultraviolet divergences 
and infinite renormalization.  The hope was disappointed, but QCD 
was found nevertheless.  A similar hope is alive for quantum 
gravity, but this time the perspectives look better.  
Perturbative string theory is (almost certainly) finite order by 
order, and loop quantum gravity reveals a discrete structure of 
space at the Planck scale which, literally, ``leaves no space'' 
for the ultraviolet divergences.

\end{description}

\section{Conclusion}

We have at least two well developed, although still incomplete, 
theories of quantum spacetime, string theory and loop quantum 
gravity.  Both theories provide a physical picture and some 
detailed results on Planck scale physics.  The two pictures are 
quite different, in part reflecting the diverse cultures from 
which they originated, high energy physics and relativity.  In 
addition, a number of promising fresh ideas and fresh approaches 
have recently appeared, most notably noncommutative geometry.  
The main physical results on quantum spacetime obtained in he 
last three years within these theories are the following.

\begin{itemize}

\item A striking result is the explicit computation of the quanta 
of geometry, namely the discrete spectra of area and volume, 
obtained in loop quantum gravity. 

\item Substantial progress in understanding black hole entropy 
has been achieved in string theory, in loop quantum gravity, and 
using other techniques. 

\item Two cosmological applications of quantum gravity have been 
proposed.  String cosmology might yield predictions on the 
spectrum of the background gravitational radiation.  According to 
Woodard and Tasmis, two-loops quantum gravity effects might be 
relevant in some cosmological models. 

\end{itemize}

Among the most serious open problems are the following. 

\begin{itemize}

\item Black hole entropy has been discussed using a variety 
of different approaches, and the relation between the 
various ideas is unclear.  What is needed in black hole 
thermodynamics is a critical comparison between the many 
existing ideas about the source of BH entropy, and possibly 
a synthesis.

\item In string theory, the key problem in view of the 
description of quantum spacetime is to find the 
background-independent formulation of the theory.

\item In loop quantum gravity, the main problem is to understand 
the low energy limit and to single out the correct version of the 
hamiltonian constraint.  A promising direction in this regard 
might be given by the spin foam models.

\item In noncommutative geometry, the problem that probably needs 
to be understood is the relation between the noncommutative 
structure of spacetime and the quantum field theoretical aspects 
of the theory.  In particular, how is renormalization affected by 
the spacetime noncommutativity?

\end{itemize}

The relations between various approaches may be closer than 
expected.  I have already pointed out some noncommutative 
geometry aspects of string theory.  String theory and loop 
quantum gravity are remarkably complementary in their successes, 
and one may speculate that they could merge or that some 
technique could be transferred from one to the other.  In 
particular, loop quantum gravity is successful in 
dealing with the nonperturbative background-independent 
description of quantum spacetime, which is precisely what is 
missing in string theory, and loops might provide 
some tools to strings.  A loop, of course, is not a very 
different object from a string.  String theory can be formulated 
as a sum over world-sheets, and loop quantum gravity can be 
formulated as a sum over surfaces.  The world sheets of string 
theory do not branch and are defined over a metric space.  In 
particular, a displaced world-sheet is a distinct world-sheet.  
The surfaces of loop quantum gravity, on the other hand, branch, 
and are defined in a background independent manner over a space 
without metric, where only their topology, not their location, 
matters.  The existence of some connection between the two 
pictures has been advocated in particular by Smolin \cite{lee}; 
and Ashoke Sen has recently introduced a notion of ``string 
networks'' into string theory, paralleling the step from loops to 
spin network in loop quantum gravity \cite{ashoke}.  \footnote{In 
``Blue Mars'', the last novel of the science-fiction Mars 
trilogy by Kim Stanley Robinson \cite{Mars}, the fundamental 
physics of the 23rd century is based on a merging between loop 
and string theories!}

\vskip1cm

In conclusion, I believe that string theory and loop quantum 
gravity do represent real progress.  With respect to few years 
ago, we now do better understand what may cause black hole 
entropy, and what a quantized spacetime might be.

However, in my opinion it is a serious mistake to claim that this 
is knowledge we have acquired about nature.  Contrary to what is 
too often claimed even to the large public, perhaps with damage 
to the credibility of the entire theoretical community, these are 
only very {\em tentative\/} theories, without, so far, a single 
piece of experiment support.  For what we really know, they could 
be right or entirely wrong.  What we really know at the 
fundamental physical level is only the standard model and general 
relativity, which, within their domains of validity have received 
continuous and spectacular experimental corroboration, month 
after month, in the last decades.  The rest is, for the moment, 
tentative and speculative searching.

But is worthwhile, beautiful, fascinating searching, which might 
lead us to the next level of understanding nature.

\end{document}